\documentclass[acmsmall,screen,nonacm,language=english,xcolor=table]{acmart}

\usepackage{amsmath}
\usepackage{bm}
\usepackage{todonotes}
\usepackage{booktabs}
\usepackage{threeparttable}
\usepackage{adjustbox}
\usepackage{longtable}
\usepackage{caption}
\usepackage{subcaption}
\usepackage{graphicx}
\usepackage{cleveref}
\usepackage{comment}
\usepackage{xurl}
\usepackage{array}
\usepackage{pifont}
\usepackage{multirow}
\usepackage{fontawesome5}
\usepackage{soul}
\sethlcolor{yellow} %
\usepackage{color}
\usepackage{colortbl} %
\usepackage{xparse}
\usepackage{xstring}
\usepackage{etoolbox}
\usepackage{xspace} %

\soulregister{\cite}{7}
\soulregister{\citet}{7}
\soulregister{\citep}{7}
\soulregister{\ref}{7}
\soulregister{\cref}{7}
\soulregister{\paragraph}{7}
\soulregister{\edit}{7}
\soulregister{\url}{7}
\soulregister{\textit}{7}
\soulregister{\textbf}{7}
\soulregister{~}{7}

\usepackage[most]{tcolorbox}
\usepackage{tikz}
\usetikzlibrary{positioning}

\newtcolorbox{sectiongoal}{
    colback=blue!5!white,    %
    colframe=blue!75!black,  %
    arc=4pt,                 %
    boxrule=0.8pt,           %
    left=10pt, right=10pt, top=8pt, bottom=8pt,
    fontupper=\small\itshape\sffamily, %
    before skip=10pt, after skip=10pt,
    enhanced,                %
    breakable                %
}

\newcommand{\opfe}{OPFE\xspace}
\newcommand{\opfes}{OPFEs\xspace}

\newcommand{\atproto}{AT Protocol\xspace}
\newcommand{\bsky}{Bluesky\xspace}

\newcommand{\example}[2]{
    \begin{tcolorbox}[sharp corners, colback=white, colframe=black!70, title={\small #1}]
    {\small #2}
    \end{tcolorbox}
}

\definecolor{greenfill}{HTML}{1b9e77}
\definecolor{greenborder}{HTML}{33a02c}
\definecolor{orangefill}{HTML}{ffA500}
\definecolor{orangeborder}{HTML}{ff7f00}
\definecolor{redfill}{HTML}{e41a1c}
\definecolor{redborder}{HTML}{e31a1c}

\definecolor{opfegrey}{HTML}{F2F2F2}

\definecolor{lhdrbg}{HTML}{CCCCCC}
\definecolor{l1fill}{HTML}{E6F2FF}
\definecolor{l2fill}{HTML}{CCE5FF}
\definecolor{l3fill}{HTML}{B3D9FF}
\definecolor{l4fill}{HTML}{99CCFF}

\newcommand{\tabhigh}{
\tikz[baseline=0.65ex]{
    \filldraw[
        fill=greenfill,
        draw=greenfill,
        line width=0.4pt
    ] (0,0) rectangle (2.6ex,2.6ex);}}

\newcommand{\high}{
\tikz[baseline=0.4ex]{
    \filldraw[
        fill=greenfill,
        draw=greenfill,
        line width=0.5pt
    ] (0,0) rectangle (2.2ex,2.2ex);}}
\newcommand{\med}{
\tikz[baseline=0.2ex]{
    \filldraw[
        fill=orangefill,
        draw=orangefill,
        line width=0.4pt
    ] (0,0) rectangle (1.6ex,1.6ex);}}
\newcommand{\low}{
\tikz[baseline=-0.3ex]{
    \filldraw[
        fill=redfill,
        draw=redfill,
        line width=0.4pt
    ] (0,0) rectangle (0.7ex,0.7ex);}}

\newcommand{\propdown}{\textcolor{red}{$\downarrow$}}
\newcommand{\propup}{\textcolor{greenfill}{$\uparrow$}}

\newcommand{\propupB}{\textcolor{greenfill}{$\pmb{\pmb{\uparrow}}$}}

\newcommand{\proph}[1]{\textit{#1}}
\NewDocumentCommand{\prop}{m o s}{%
  \leavevmode
  \proph{%
    \IfValueTF{#2}{%
      \begingroup
      \setlength{\fboxsep}{0pt}%
      \smash{\propScore{#2}} #1%
      \endgroup
      \kern-0.1pt 
    }{#1}%
  }%
}

\newcommand{\propScore}[1]{%
  \IfEqCase{#1}{%
    {high}{\high~high}%
    {High}{\high~High}%
    {medium}{\med~medium}%
    {Medium}{\med~Medium}%
    {low}{\low~low}%
    {Low}{\low~Low}%
  }[#1]%
}

\newcommand{\appr}[1]{#1}

\title[Open Platform Field Experiments]{Open Platform Field Experiments: Expanding the Design Space of Experimental Research on Social Media}

\author{Jordi Guillem Condom-Tibau}
\authornote{Corresponding author.}
\affiliation{%
  \institution{University of Pisa}
  \city{Pisa}
  \country{Italy}}
\affiliation{%
  \institution{Institute for Informatics and Telematics, National Research Council (IIT-CNR)}
  \city{Pisa}
  \country{Italy}}
\email{jordi.condom@phd.unipi.it}

\author{Giovanni Puccetti}
\affiliation{%
  \institution{Institute of Science and Technologies of Information “A. Faedo”, National Research Council (ISTI-CNR)}
  \city{Pisa}
  \country{Italy}}
\email{giovanni.puccetti@isti.cnr.it}

\author{Clara Bacciu}
\affiliation{%
  \institution{Institute for Informatics and Telematics, National Research Council (IIT-CNR)}
  \city{Pisa}
  \country{Italy}}
\email{clara.bacciu@iit.cnr.it}

\author{Matteo Abrate}
\affiliation{%
  \institution{Institute for Informatics and Telematics, National Research Council (IIT-CNR)}
  \city{Pisa}
  \country{Italy}}
\email{matteo.abrate@iit.cnr.it}

\author{Stefano Cresci}
\affiliation{%
  \institution{Institute for Informatics and Telematics, National Research Council (IIT-CNR)}
  \city{Pisa}
  \country{Italy}}
\email{stefano.cresci@iit.cnr.it}

\renewcommand{\shortauthors}{Condom-Tibau et al.}

\begin{document}

\begin{abstract}
Despite a growing demand for causal evidence about social media, independent researchers remain severely constrained in their ability to conduct experiments directly on online platforms. To cope, multiple methodological workarounds have emerged---from controlled surveys and simulations to client-side overlays and platform partnerships---each requiring distinct trade-offs between desirable experimental properties. The recent emergence of open social media platforms offers a qualitatively different methodological opportunity. Here we propose a design space of social media experimentation and discuss \textit{Open Platform Field Experiments} (\opfes{}). OPFEs represent a distinct class of experimental approaches that enable independent researchers to directly intervene on functional platform components---such as clients, recommendation systems, and moderation services---within live social media environments. Through a comparative analysis of experimental archetypes, we show that OPFEs occupy a previously unexplored region of the design space. We then bridge theory and practice by characterizing the architectural and governance elements that enable OPFEs, mapping them onto \bsky{} and the \atproto{}, and illustrating the end-to-end lifecycle of a complete \opfe{} design. Overall, this work establishes OPFEs as a practical methodological paradigm for independent, transparent, and ecologically grounded experimentation on open social media.
\end{abstract}




\maketitle

\section{Introduction}
\label{sec:intro}
A select few privately governed digital platforms control the information environments of billions of individuals.
As these large online spaces shape what people read and with whom they interact at a societal scale~\citep{ibrahim2026systematic}, a growing range of scholars---from the computer and social sciences, psychology, communication, economics, and policy---have converged on a shared methodological need: the ability to generate causal evidence about how platforms influence behavior and downstream societal outcomes~\citep{bak2021stewardship}. Questions about how ranking algorithms, content moderation policies, or AI-generated outputs influence user behavior have moved from academic inquiry to matters of public consequence~\citep{costello2024durably,munger2025did,hackenburg2025levers}, now entangled with geopolitical tensions over platform governance and technological sovereignty~\citep{burwell2026digital}. More broadly, online platforms are increasingly used not only as objects of experimental inquiry, but also as infrastructures through which a wide range of social, political, economic, and behavioral phenomena can be experimentally studied~\citep{guess2021experiments,bond201261,athey2023digital}. Early work in these domains relied predominantly on observational and descriptive analyses~\citep{mosleh2022field,guo2025estimating}, but such approaches are fundamentally limited in their ability to disentangle causal mechanisms from mere correlation~\citep{lorenz2023systematic}. Particularly so in environments characterized by complex feedback loops, algorithmic curation, and network effects~\citep{pedreschi2025human}. In contrast, experimental methods enable controlled manipulation of exposure and intervention, and offer a uniquely powerful lens to identify causal effects. 

The ability to conduct field experiments on large social media platforms\footnote{While the term \textit{platform} encompasses a broad spectrum of digital ecosystems, any reference to platforms throughout this paper refers strictly to social media platforms.} has historically been tightly constrained by their status as walled gardens: privately governed systems with limited external access. These constraints operate along multiple dimensions. \textit{Infrastructure opacity} limits researchers’ ability to intervene on core functionalities such as interface, recommendation, or moderation. It reflects the difficulty to \textit{change} the system. \textit{Data opacity} restricts access to the behavioral traces and exposure histories necessary to design and evaluate interventions. It reflects the difficulty to \textit{observe} the system. \textit{Epistemic opacity} limits what is known about how platform processes generate observed outcomes. Beyond restricted data access, this includes the lack of transparency around algorithmic configurations, training data, system dynamics, and internal organization~\citep{ruths2014social}. It thus reflects the difficulty to \textit{know} the system. Among the unknowns is also the design and deployment of platform-run experiments. In fact, while platforms themselves routinely run large-scale experiments, these are typically undisclosed, methodologically opaque, and non-reproducible~\citep{polonioli2023ethics}. In a small number of cases, platforms have collaborated with selected academics to enable controlled experiments~\citep{gonzalez2023asymmetric,guessResharesSocialMedia2023,guessHowSocialMedia2023,nyhan2023like,allcott2024effects}, but such arrangements are rare, ad hoc, and highly selective~\citep{turner2025we}. In other words, they are fundamentally restricted \textit{by design}---as a handful of private actors cannot accommodate the experimental needs of a global research and policy community through discretionary and case-by-case collaboration. The result is a persistent asymmetry in experimental capacity. Platforms possess extensive capabilities to test and optimize, often in service of narrow commercial objectives~\citep{turner2025we}, while independent researchers, policymakers, and civil society actors have a multitude of open questions but limited capacity to generate causal evidence about them. 

In response to these constraints, scholars across disciplines have developed a diverse set of strategies to approach experimental studies on social media. Some rely on surveys administered in controlled laboratory settings or construct simplified representations of platform environments to manipulate exposure and measure responses with extensive experimental control~\citep{vragaNewsLiteracyMessages2022,sharevskiSoftModeration2022}. Others turn to computational simulations, now increasingly powered by generative AI~\citep{gao2024large}, to model user behavior and platform dynamics at scale~\citep{muric2022large,truong2024quantifying}. Others still attempt to intervene directly on live platforms, for instance by means of automated accounts~\citep{bilewicz2021artificial}, or develop ad hoc technical tools such as browser extensions that modify what the user sees locally~\citep{piccardi2025reranking}. While each of these approaches has proven valuable for specific research goals, they inevitably involve trade-offs that limit their ability to generate robust causal evidence. Highly controlled settings offer precision but lack ecological validity~\citep{mosleh2022field}. Simulations scale easily but depend on strong and often untestable assumptions~\citep{rossi2024problems,ye2026stop}. In situ interventions can capture real behavior but may raise ethical, legal, and methodological concerns, or be constrained in scope, scalability, and generalization~\citep{straub2024public}. As a result, different communities have adopted different methodological compromises, reflecting their respective priorities and constraints. Taken together, this picture underscores a common limitation: in the absence of direct access to platform infrastructures, researchers have been forced to obtain causal evidence on social media either by abstracting away its complexity, by artificially recreating it, or by engaging with proxies rather than the underlying mechanisms.

Recent developments in decentralized and open social media introduce a qualitatively different set of possibilities for experimental research. In contrast to traditional platforms, these systems expose multiple layers of open functionality---at the level of protocols, algorithms, clients, and data---creating an ecosystem in which some key components of the environment can be accessed, modified, or recombined by external actors~\citep{kleppmannBlueskyProtocolUsable2024}. This form of openness does not eliminate constraints, but it shifts them away from the bottleneck imposed by closed platforms toward a scenario where challenges can be engaged with more directly. Rather than relying on discretionary, case-by-case collaboration with platform providers, researchers can design and deploy interventions by leveraging the system’s built-in components. Early work has begun to explore these possibilities, suggesting the emergence of a class of \textit{open-platform field experiments} (OPFEs) that operate directly on elements of real-world social media systems without requiring full platform control~\citep{popowski2026social,el2026bonsai,chujo2026exploring}. Compared to prior approaches, this setting reduces the need to abstract away complexity, simulate platform dynamics, or rely on proxies, enabling more direct engagement with mechanisms that shape user experience and behavior. At the same time, these opportunities come with their own challenges, including the need to understand evolving technical architectures and account for new forms of partial control and uncertainty. Open platforms might not fully resolve the tensions inherent in social media experimentation, but they expand the design space for experiments on social media in ways that warrant systematic examination.

\begin{table}[t]
    \centering
    \small
    \caption{Reading guide to the core contributions of this study. For each subsection, we summarize the main object discussed, highlighted in \textit{italics}, allowing readers to quickly identify and navigate to the core topics.}
    \begin{tabular}{lp{0.89\linewidth}@{}}
        \toprule
        \textbf{section} & \textbf{description} \\ 
        \midrule
        \ref{sec:design-space} & Design space of social media experiments \\
        \quad\ref{sec:design-space-properties} & Definition of the \textit{desirable properties} through which experimental approaches are characterized \\
        \quad\ref{sec:design-space-approaches} & Definition of a typology of social media \textit{experimental archetypes} \\
        \quad\ref{sec:design-space-evaluation} & \textit{Comparative analysis} of the experimental archetypes in terms of the desirable properties \\ 
        \midrule
        \ref{sec:opfes} & Open-platform field experiments (OPFEs) \\
        \quad\ref{sec:opfes-design-space} & Definition of OPFEs and their distinctive \textit{experimental capabilities} \\
        \quad\ref{sec:opfes-layers} & Description of the \textit{open platform elements} and components that enable OPFEs \\ 
        \midrule
        \ref{sec:bluesky} & OPFEs on Bluesky \\
        \quad\ref{sec:bluesky-elements} & Description of the \textit{Bluesky and AT Protocol infrastructure} relevant to OPFEs \\
        \quad\ref{sec:bluesky-example} & \textit{Concrete design} of an illustrative OPFE on Bluesky \\
        \bottomrule
    \end{tabular}
    \label{tab:reading-guide}
\end{table}

\paragraph{Contributions} Against this background, we investigate how \opfes{} reshape the design space of social media experimentation. To reach this overarching goal, we make four contributions. First, we propose a design space that systematizes the main archetypes of social media experiments and identifies a set of desirable properties through which their experimental capabilities can be characterized. Such a framework is currently lacking in literature, yet provides the common conceptual space necessary to systematically position otherwise heterogeneous approaches. Second, we comparatively evaluate the archetypes along the identified dimensions, showing that \opfes{} occupy a distinct region of the design space. Third, we characterize the open platform elements and components that enable \opfes{}, explaining how their technical and governance designs give rise to their distinctive methodological profile. Finally, we demonstrate how \opfes{} can be realized on \bsky{}, mapping the conceptual model onto a concrete open social media ecosystem. Together, these contributions progress from a general theory of the experimental design space to its concrete operationalization within an existing open social media ecosystem.

\paragraph{Structure and Organization} The organization of these contributions is summarized in Table~\ref{tab:reading-guide} and briefly reported in the following. Section~\ref{sec:design-space} defines the design space of social media experimentation and comparatively positions existing archetypes in the space. Section~\ref{sec:opfes} introduces \opfes{} and motivates their properties in terms of their constituting elements. Section~\ref{sec:bluesky} presents elements of \bsky{} and the \atproto{} that are relevant for \opfes{}, and shows an end-to-end implementation blueprint for an \opfe{} on \bsky{}. Beyond our core contributions, Section~\ref{sec:rel-work} discusses the existing literature in terms of open and decentralized platforms and social media experiments. Section~\ref{sec:methodology} presents our methodology in detail. Finally, Section~\ref{sec:discussion} discusses the implications of our work while Section~\ref{sec:conclusions} draws the conclusions.

\paragraph{Methodology} We develop the proposed design space through an iterative hybrid deductive-inductive process. Deductively, we identify candidate experimental properties from theoretically motivated methodological requirements, while inductively refining both the identified experimental archetypes and properties through representative examples drawn from the literature on social media experimentation. We then comparatively assess the resulting families of experimental archetypes along the identified properties using a coarse-grained three-level ordinal scale, providing a systematic yet tractable characterization of their relative capabilities and trade-offs.

\paragraph{Significance} This work has both conceptual and practical implications. By structuring the design space, it provides a common vocabulary through which existing, emerging, and hybrid experimental approaches can be systematically positioned and compared, bringing structure to a methodologically rich but diverse and fragmented landscape. For researchers and practitioners, the framework can inform methodological choices by clarifying which approaches best match a given research question, set of experimental requirements, and available resources, or conversely, what can reasonably be achieved when a particular approach is chosen. The characterization of \opfes{} connects experimental capabilities to concrete properties of open platforms, identifying which architectural and governance choices enable different forms of experimentation. This connection can inform the design of future research-friendly platforms and infrastructures, particularly amid growing concerns around technological sovereignty, privacy, and independent scrutiny of digital systems. Finally, the instantiation on \bsky{} translates these abstractions into an actionable example, providing researchers with a practical reference for concretely conducting \opfes{}.

\section{Related Work}
\label{sec:rel-work}
We first provide background on open and decentralized social media platforms, while reviewing the growing body of research that studies them. We then turn to experimental approaches to social media research by surveying its vast methodological landscape.

\subsection{Open and Decentralized Social Media}
\label{sec:rel-work-open-social-media}
Although often discussed together, \textit{openness} and \textit{decentralization} refer to distinct properties of online platforms. Decentralization concerns the distribution of technical infrastructure and decision-making authority (e.g., governance) across multiple actors, whereas openness refers to platform components that are externally accessible and modifiable. While these properties frequently co-occur, they are neither equivalent nor mutually dependent. Hereafter we mainly focus on open platforms, because openness directly determines the extent to which researchers can independently access, modify, and intervene on platform components. Decentralization may facilitate or reinforce such opportunities, but it is not, by itself, sufficient to enable them.

Rather than operating as monolithic services, open platforms are organized around modular architectures in which different components can be independently accessed, replaced, or extended. Typical such components are protocols, data interfaces, clients, ranking systems, and moderation mechanisms. A common design pattern across open platforms is the separation between a protocol layer, which defines rules for identity, data exchange, and network coordination, and an application layer, which provides user-facing services such as clients, feeds, or moderation tools. This architectural openness creates opportunities for external actors to inspect, modify, and build upon parts of the platform infrastructure. In practice, these ideas have been realized through a variety of technical architectures. Early systems are predominantly federated networks, such as \textit{Mastodon},\footnote{\url{https://joinmastodon.org/}} which is part of the broader Fediverse built on the ActivityPub\footnote{\url{https://www.w3.org/TR/2018/REC-activitypub-20180123/}} standard~\citep{la2021understanding}. Other ecosystems adopt protocol-centric designs that further decouple identity, content storage, and user interfaces. Examples include the Secure Scuttlebutt (SSB) protocol,\footnote{\url{https://ssbc.github.io/scuttlebutt-protocol-guide/}} used by applications such as \textit{Manyverse},\footnote{\url{https://www.manyver.se/}} and \textit{Nostr},\footnote{\url{https://nostr.com/}} which relies on relay-based communication and cryptographic identity. More recent platforms such as \textit{Bluesky},\footnote{\url{https://bsky.app/}} \textit{W Social},\footnote{\url{https://wsocial.news/}} and \textit{Mu Social}\footnote{\url{https://eurosky.tech/}} are built on the Authenticated Transfer Protocol (AT Protocol)~\citep{kleppmannBlueskyProtocolUsable2024}, while ecosystems such as \textit{Farcaster}\footnote{\url{https://farcaster.xyz/}} and \textit{Lens}\footnote{\url{https://lens.xyz/}} explore alternative approaches based on blockchain technologies. Although these systems differ substantially in their architecture and governance, they share a common trend toward exposing platform functionality in ways that enable external participation and innovation. Recent initiatives such as the European Social Stack\footnote{\url{https://european.social/}} further illustrate how open social platforms are increasingly framed not only as technical alternatives to closed platforms, but also as elements of digital sovereignty, democratic resilience, and public information infrastructure.

The emergence of open and decentralized platforms has attracted considerable attention from users and scholars alike. While legacy systems remain dominant, users and communities are increasingly diversifying their online presence by exploring decentralized alternatives such as Mastodon~\citep{ng2025journalists,he2023flocking,cava2023drivers} and Bluesky~\citep{jeong2023user,quelle2025academics}. At the same time, scholars have increasingly turned to these platforms because of their open data ecosystems, which offer new opportunities at a time when access to data from mainstream social media has become progressively restricted~\citep{tromble2021have,goanta2026great}. As a result, open platforms have rapidly become valuable environments for studying online behavior and social dynamics. As expected for a relatively recent technological ecosystem, most existing research has been observational and descriptive in nature. Studies have examined migration dynamics and network formation~\citep{balduf2025bootstrapping}, self-moderation and blocking behavior~\citep{bono2026self,sokoto2026open}, the impact of decentralized moderation on network structure and information spread~\citep{arregui2026effects}, the efficacy of large language models (LLMs) for automated platform moderation~\citep{chou2026open}, how community rules change across Mastodon instances~\citep{muralidharan2026federating}, Bluesky starter packs~\citep{failla2026structure}, the effects of social media consumption on psychosocial well-being~\citep{pal2026hidden}, or AI-generated content in social media~\citep{hajumpee2026ai}. More generally, open platforms have rapidly become valuable sources of behavioral data for computational social science~\citep{quelle2025bluesky,jeong2024bluetempnet,failla2024m,balduf2024looking,nogara2026longitudinal, smith2026blue, salloum2025politics}.

Beyond data access, however, researchers have also begun leveraging the openness of platform infrastructure as a means of conducting experimental research. A few recent studies have explored custom feed systems and feed personalization mechanisms~\citep{greenwood2026paper}, developed tools that allow users to construct intentional and personalized feeds~\citep{el2026bonsai}, and employed structured elicitation methods to help users articulate desired recommendation algorithms~\citep{popowski2026social}. Other work has investigated the effects of alternative feed designs during major political events~\citep{bradyRedesigningAlgorithms2026} or modified platform clients to study interventions aimed at mitigating emotional manipulation~\citep{chujo2026exploring}. This emerging literature suggests that open platforms are increasingly serving not only as sources of observational data, but also as infrastructures for field experimentation.

\subsection{Social Media Experiments}
\label{sec:rel-work-experiments}

Experimental research on social media has emerged across a wide range of disciplines. As interest in causal questions about online behavior has grown, researchers have developed a diverse set of approaches for studying social media and its effects. Initially, different disciplinary traditions gravitated toward different experimental strategies. For instance, researchers in psychology and communication often relied on surveys and mock interfaces~\citep{guess2020digital,pennycookShiftingAttention2021,vragaNewsLiteracyMessages2022}, whereas computer scientists more frequently developed computational simulations, browser extensions, and custom software tools~\citep{fidone2026evaluating,piccardi2025reranking}. The multitude of existing approaches reflects differences in methodological traditions, technical capabilities, and degrees of access to platform infrastructures more than differences in the underlying phenomena being studied. Indeed, although multiple communities sought answers to similar questions, they have often relied on different experimental approaches. As the field matures, however, these boundaries are becoming increasingly blurred. As a result, the literature encompasses an increasingly heterogeneous collection of experimental strategies that vary substantially in their realism, complexity, and scope.

Some experimental approaches leverage affordances that are already available within existing platforms. Researchers have used direct messages, advertisements, moderation tools, community management features, and other native functionalities to deliver interventions without modifying the underlying platform. Many such studies treat social media platforms themselves as the object of inquiry, using experiments to evaluate moderation and governance mechanisms~\citep{jhaver2019did,matias2019preventing,yildirimSuspensionWarnings2023}, the spread of misinformation and corrective interventions~\citep{pennycookShiftingAttention2021}, or other platform-mediated behaviors and outcomes. For instance, researchers have used native reward systems, such as Reddit awards, to conduct field experiments comparing human susceptibility to social influence from artificial agents versus human peers~\citep{oda2026field}. At the same time, social media platforms are increasingly used not merely as objects of study, but as infrastructures through which experiments on broader social phenomena can be conducted. Their large user populations, rich communication environments, and ability to deliver randomized interventions at scale make them attractive settings for studying behaviors that extend well beyond the platforms themselves. Researchers have therefore used social media to investigate political mobilization~\citep{aggarwal20232}, civic engagement~\citep{bond201261}, public health communication~\citep{athey2023digital,donati2024facebook}, and related societal outcomes. In these cases, social media functions primarily as a vehicle for recruitment, exposure, and measurement, enabling online experiments about offline social and behavioral processes.

Researchers have also developed external tools that modify users' experience of a platform without requiring cooperation from platform providers. These include browser extensions that alter the presentation or ranking of content, modified clients that provide alternative interfaces to the same underlying platform, and software systems that interact with platform APIs to deploy interventions or manage online communities. Such approaches have been used to introduce credibility prompts~\citep{bhuiyanFeedReflect2018}, re-rank information feeds~\citep{piccardi2025reranking}, and support community-led moderation and policy evaluation~\citep{matias2018civilservant}. More generally, they enable forms of experimentation that go beyond the affordances intentionally exposed by platforms while remaining deployable by independent researchers.

Others recreated parts of social media environments in controlled settings. These approaches range from static mock interfaces and survey-based experiments to dedicated platforms designed to reproduce selected components of social media. Such environments have been used to study content moderation decisions and bystander intervention in cyberbullying~\citep{taylor2019accountability}, misinformation susceptibility and news literacy~\citep{micallefFakey2021}, feed composition and context effects~\citep{roggenkamp2025dice}, linguistic media bias~\citep{hinterreiterNewsUnfold2025}, discourse civility~\citep{seering2019designing}, and feedback exchange~\citep{wu2021better} under carefully controlled conditions. Related approaches include computational simulations that model user behavior and platform dynamics through synthetic models and populations, allowing researchers to conduct experiments that would be infeasible or prohibitively costly in real systems. Within this computational paradigm, agent-based models (ABMs) have been used to study filter bubbles~\citep{geschke2019triple}, information cascades~\citep{franken2021cascades}, how algorithmic bias amplifies opinion polarization~\citep{sirbu2019algorithmic} and echo chambers~\citep{baumann2020modeling}. A recent and rapidly growing strand of work increasingly employs generative AI agents to simulate individual users and their interactions on social platforms. These approaches have been used in literature to study content moderation strategies~\citep{fidone2026evaluating}, to evaluate feed algorithms~\citep{tornberg2023simulating}, to simulate opinion dynamics~\citep{chuang2024simulating}, and to create a digital twin of a social media platform~\citep{rossetti2026social}.  While these approaches offer unprecedented flexibility and scale, they also require researchers to specify a large number of assumptions about user behavior and platform dynamics. Their validity therefore hinges on how faithfully these assumptions mirror real-world processes~\citep{taday2026assessing}. This challenge is particularly pronounced in generative-agent simulations, where agent behavior emerges from highly complex foundation models containing billions of parameters and internal decision processes that remain only partially understood~\citep{lin2026illusion}.

Finally, at the opposite end of the spectrum are experiments conducted in direct collaboration with platform providers. Large online platforms routinely run experiments to optimize user engagement, content consumption, advertising performance, and other operational or economic objectives~\citep{bond201261}. More rarely, however, these experimental capabilities are made available to academic researchers. The most notable such example is the research partnership to understand Facebook and Instagram’s role in the US 2020 election,\footnote{\url{https://research.facebook.com/2020-election-research/}} a large-scale collaboration between Meta and independent scholars that enabled a series of experiments on real platform users. The resulting studies investigated the effects of feed ranking algorithms on political attitudes and polarization~\citep{guessHowSocialMedia2023}, content resharing on information diffusion and political beliefs~\citep{guessResharesSocialMedia2023}, and ideological segregation through exposure to like-minded sources~\citep{nyhan2023like}. Other examples include collaborations with Twitter/X to study interventions aimed at reducing offensive content and improving the quality of online discourse~\citep{katsarosReconsideringTweetsIntervening2022}, and partnerships with Reddit to evaluate proactive, community-specific moderation tools~\citep{horta2025post}. Together, these efforts have produced some of the largest and most consequential social media experiments conducted to date, while also illustrating the opportunities and dependencies associated with platform-mediated research.

\section{Methodology}
\label{sec:methodology}
\paragraph{Scope.} Our overarching goal is to investigate how \opfes{} reshape the design space of social media experimentation and to understand how they compare to existing experimental approaches. To this end, throughout the work we adopt the perspective of researchers and practitioners conducting social media experiments. Our analysis thus focuses on the capabilities, constraints, and trade-offs that different experimental approaches expose to experimenters. Other important perspectives, such as participants' experience, burden, or welfare are outside the scope of the present work. This restriction allows us to define a focused and tractable framework tailored to the methodological question addressed in this paper.

\paragraph{Design Space Development.} A prerequisite for studying \opfes{} is the availability of a common framework within which different experimental archetypes can be systematically positioned and compared. Existing work does not provide a unified formalization of this design space~\citep{lorenz2023systematic,mosleh2022field}, so we first constructed one that captures the principal approaches to social media experimentation together with the properties through which they can be meaningfully compared. This design space is therefore not an end in itself, but the methodological instrument that enables the subsequent comparative analysis of \opfes{}~\citep{maclean2020questions}. It was developed through an iterative hybrid deductive-inductive process~\citep{dubois2002systematic}. Deductively, we began from theoretically motivated expectations about the methodological requirements of social media experimentation, identifying candidate properties that we considered fundamental for experimental research, including the ability to implement interventions, observe outcomes, recruit participants, ensure compliance, and preserve ecological validity. Inductively, we repeatedly confronted these preliminary dimensions with the existing literature on social media experimentation, examining representative experiment archetypes to identify recurring capabilities, limitations, trade-offs, and meaningful distinctions not adequately captured by the current formalization. As done in recent design space studies~\citep{zhang2024form}, we adopted purposeful sampling~\citep{palinkas2015purposeful} in place of an exhaustive literature review. We thus selected representative examples from the principal communities conducting social media experiments, including CSCW, CHI, computational social science, human behavior research, psychology, communication, and related disciplines. The cited literature thus contains illustrative examples of the identified experimental approaches, rather than being an exhaustive catalog of all relevant work.

Throughout the process, we deliberately balanced expressivity with parsimony. We restricted the analysis to the principal families of experimental archetypes, without explicitly modeling hybrid or composite approaches, as these would substantially increase the complexity of the design space while providing limited additional insight for the objectives of this study. Likewise, we selected only those properties that meaningfully distinguish among the identified approaches and are necessary to characterize their experimental capabilities. Other potentially relevant aspects of social media experimentation, such as the specific research questions motivating a study, were intentionally left outside the scope of the framework. Although such factors undoubtedly influence methodological choices, modeling them would substantially broaden the scope of the present work and address a different research problem.

\paragraph{Comparative Assessment.} Once the design space had been defined, we comparatively assessed each experimental archetype with respect to the identified properties, creating a methodological profile for each. Specifically, each property captures the extent to which an approach supports an ideal experimental condition and how easily researchers can achieve that condition in practice. A\protect\high~high score indicates that the corresponding property is consistently achieved with low effort or intrinsically supported; a\protect\med~medium score indicates that it can be obtained only partially or with technical effort, methodological compromises, or specific mitigation strategies; a\protect\low~low score indicates that support is generally limited, requiring substantial additional effort or being infeasible in many practical settings. The scores were assigned through comparative expert assessment by the authors, informed by the literature examined during framework development and by the defining characteristics of each family of experimental approaches. The objective of this assessment is comparative rather than evaluative: higher scores indicate stronger support for a specific property, not the overall superiority of an experimental approach. We intentionally employ a coarse-grained three-level ordinal scale because substantial methodological heterogeneity exists both across and within the considered families of approaches. Consequently, the assigned scores should be interpreted as comparative characterizations of broad methodological tendencies rather than precise or absolute measurements.

\paragraph{Iterative Refinement.} Design space development and comparative assessment proceeded through multiple rounds of refinement over approximately four months. Following each iteration, the identified approaches, properties, and their definitions were jointly reviewed by all authors through open discussion. Once a sufficiently stable design space had emerged, we performed the comparative assessment of the approaches using the identified properties. The resulting scores were also discussed collectively, with disagreements prompting further refinement of the property definitions. We then repeated the comparative assessment using the revised framework. This iterative process continued until successive iterations no longer produced substantive changes to either the framework or the comparative evaluations, and all authors reached consensus on both the resulting design space and the assigned scores.

\section{The Design Space of Social Media Experiments}
\label{sec:design-space}
Researchers design social media experiments with distinctive research questions in mind while working around various constraints imposed by platforms. This has led to the development of a variety of experimental approaches, each addressing the needs of social media experiments in different ways~\cite{warburton2026conduct}. This section proposes a systematic comparison of these strategies. First, we characterize social media experiments in terms of the main \emph{desirable properties} that an experimental approach should ideally possess. Afterwards, we organize these diverse approaches into a \emph{typology of existing experiment archetypes} and assess their respective strengths and weaknesses in terms of the desirable properties.

This systematization leads us to a working definition of a \emph{design space} for social media experiments.

\subsection{Desirable Properties}
\label{sec:design-space-properties}
Each of the following properties captures the degree to which an approach supports an experimental condition, as well as how easy it is for researchers to achieve such a condition in practice.
For clarity, properties and their scores are \proph{italicized} throughout the text.

\subsubsection{Ecological Validity}
Social media experiments are intended to inform on what happens on real platforms, and researchers value studies that closely resemble the environment they want to investigate. \prop{Ecological validity} (EV) captures the extent to which the experimental setting reflects real-world platform environments and user behavior. An experimental approach with \prop{ecological validity}[high] implies that its findings are more likely to generalize to actual social media contexts. It also encompasses outcome validity, whether the measured outcomes correspond to the theoretical constructs of interest. For example, whether observed behavior is measured directly or approximated through self-reported intentions or beliefs. In addition, this property entails population validity, or the extent to which the participants involved in the experiment are representative of the broader user population of interest. Approaches that abstract away from these dimensions may offer greater control~\citep{warburton2026conduct}, but risk studying behavior, outcomes, or populations that do not fully correspond to real-world conditions, resulting in \prop{ecological validity}[low].

\subsubsection{Controllability}
Social media experiments address a broad range of research questions, and each is answerable only if relevant \textit{mechanisms} of the platform can be manipulated. These mechanisms shape how users interact with the platform, either a real one or a proxy, and include elements of the user interface, as well as content exposure, ranking, or moderation processes.
\prop{Controllability} (Ct) refers to the extent to which researchers can manipulate such mechanisms. This includes not only the ability to define an intervention, but also to ensure that users actually receive the intended treatment (i.e., control over exposure). It also encompasses the capacity to implement and enforce randomization at different levels (e.g., user, post, feed, community, or algorithmic level), which is essential for robust causal inference. \prop{controllability}[High] enables precise experimental interventions and clearer identification of causal effects. However, in many settings---particularly on closed platforms---control is limited to predefined actions and mediated by platform systems, constraining both how treatments can be delivered and how reliably they can be assigned, ultimately resulting in \prop{controllability}[low].

\subsubsection{Observability}
An experiment on social media must collect relevant behavioral, exposure, and outcome data. This includes not only observable user actions (e.g., clicks, shares, comments), but also information about what users were exposed to and under what conditions.
\prop{Observability} (Ob) captures the extent to which such data can be accessed and reliably measured. It also entails the distinction between public and non-public data. While some approaches only allow access to publicly visible interactions, others may provide visibility also into private or restricted activities (e.g., direct messages, passive consumption, unpublished drafts, login times). This property thus also captures the ability to track volatile or dynamic changes in the system, such as edits, deletions, moderation labels, or account migrations, which can affect both measurement and interpretation. \prop{observability}[Low] can obscure causal mechanisms and introduce measurement bias, particularly when key variables such as algorithmic exposure or non-public interactions are not directly accessible. Approaches vary widely in the granularity, completeness, and reliability of the data they can provide.

\subsubsection{Compliance}
\prop{Compliance} (Cp) denotes the extent to which experiments can be conducted in accordance with applicable ethical standards as well as legal and platform regulations---including informed consent, minimal harm, transparency, and data protection. A key aspect of ethical compliance concerns user agency and consent granularity, including whether users can knowingly opt into experimental conditions (e.g., modified feeds, clients, labeling systems, or data donation), whether they are provided with meaningful opt-out mechanisms, and whether they are adequately informed or debriefed about the nature and outcomes of the study~\citep{polonioli2023ethics}. Different experimental settings raise different ethical challenges, particularly when interventions occur without user awareness or affect real-world information environments. This property also encompasses broader security and ethical considerations, including the risk of dual use, where experimental infrastructures could be repurposed for harmful activities such as spam, manipulation, or exposure to harmful content. Legal requirements further shape what constitutes compliant experimentation and vary across jurisdictions. For example, regulations such as the General Data Protection Regulation\footnote{\url{https://eur-lex.europa.eu/eli/reg/2016/679/oj/eng}} (GDPR) in the European Union impose strict constraints on data processing, consent, and user rights, while other frameworks like the Digital Services Act\footnote{\url{https://eur-lex.europa.eu/eli/reg/2022/2065/oj/eng}} (DSA) and the AI Act\footnote{\url{https://eur-lex.europa.eu/eli/reg/2024/1689/oj/eng}} introduce additional obligations related to platform governance and algorithmic systems. In contrast, regulatory regimes in other contexts, such as the United States, may differ in scope and enforcement, leading to variation in what is legally permissible. As a result, the feasibility of compliant experimentation is not uniform, as it depends both on the methodological approach and on the regulatory and institutional environment in which the research is conducted~\citep{lubin2024mapping}. Finally, compliance also includes adherence to platform-specific terms of service, conditions, and API policies, which may restrict automated data collection, content manipulation, or the use of bots, even when they are otherwise legally allowed. \prop{compliance}[Low] indicates that adhering to all required regulations may be beyond researchers' agency.

\subsubsection{Platform Independence}
\prop{Platform independence} (PI) refers to the extent to which experiments can be conducted without requiring discretionary approval or direct collaboration from platform providers. \prop{platform independence}[High] allows researchers to design and deploy studies autonomously, define research questions more freely, and make experimental approaches more broadly accessible to the research community. In contrast, approaches with \prop{platform independence}[low] depend on platform access or cooperation, and are thus shaped by the availability of such collaborations and possibly by platform-specific constraints and priorities. As a result, this property affects not only the feasibility of conducting experiments, but also the degree of flexibility researchers have in designing their studies.

\subsubsection{Scalability}
\prop{Scalability} (Sc) refers to the extent to which an experiment can be deployed to a sufficiently large and relevant population. Larger-scale experiments enable more precise estimation, greater statistical power, and the study of heterogeneous effects across user groups. However, this property often depends on access to existing user populations and infrastructure, and may come at the cost of reduced control or increased dependence on platform systems. It also interacts with ethical and legal constraints, as scaling to large populations can make it more difficult to obtain informed consent, provide meaningful opt-out mechanisms, or ensure adequate debriefing and accountability. As a result, approaches with \prop{scalability}[high], while easily scaling to large numbers of participants, may face greater challenges in meeting compliance requirements. In contrast, approaches that prioritize participants' consent and oversight may inherently lead to \prop{scalability}[low].

\subsubsection{Operational Accessibility} Social media experiments are conducted across a wide range of disciplines, by researchers whose technical backgrounds and access to engineering support vary considerably. \prop{Operational accessibility} (OA) refers to the extent to which an experimental approach can be designed, deployed, and maintained with reasonable levels of technical expertise, infrastructure, and engineering effort. Approaches with \prop{operational accessibility}[high] can be implemented using widely available tools, or require limited specialized knowledge or system integration. In contrast, some experimental settings rely on complex infrastructures, custom software, or tight integration with multiple platform components, which can increase the technical demands on researchers. While such setups may enable richer and more realistic experiments, \prop{operational accessibility}[low] raises entry barriers and can affect the feasibility and reproducibility of the research. As such, this property captures the practical ease with which different experimental approaches can be adopted and sustained by the broader research community.

\subsubsection{Experimental Isolation}
\prop{Experimental isolation} (EI) captures the degree to which the treatment assigned to one unit (e.g., a user, post, or community) is independent of other units and of concurrent or sequential interventions. \prop{experimental isolation}[Low] implies that the effects of an intervention are not isolated and therefore can be confounded by spillovers, cross-condition contamination, or interactions with other experiments. In social media settings, achieving isolation can be challenging due to network-mediated interactions (e.g., replies, reposts, social contagion) violating assumptions of unit independence and creating interference between experimental conditions~\citep{eckles2017design}, as well as system-level dynamics involving complex feedback loops such as algorithmic amplification and shared information environments~\citep{pedreschi2025human}. Dependencies may also arise from measurement processes (e.g., edits, deletions, moderation actions) or from shared infrastructures in which multiple concurrent experiments interact. \prop{experimental isolation}[High] facilitates causal identification by enabling clearer attribution of observed effects to the intended treatment. However, enforcing isolation often requires limiting or abstracting away the interdependent dynamics that characterize real-world online platforms, highlighting a fundamental trade-off between experimental control and realism.

\subsubsection{Reproducibility}
\prop{Reproducibility} (Rp) refers to the extent to which an experiment can be autonomously replicated under comparable conditions and yield consistent results. It depends on the transparency of experimental procedures, the stability of the underlying environment, and the accessibility of the necessary data and infrastructure. Approaches that rely on proprietary systems or one-off implementations may exhibit \prop{reproducibility}[low], while more open, controlled, or standardized settings typically facilitate it.

\subsection{Typology of Experimental Archetypes}
\label{sec:design-space-approaches}
Building on the desirable properties introduced above, we organize existing experimental archetypes into a typology that captures the main strategies used to generate causal evidence about social media platforms. The proposed typology is not intended to be exhaustive or rigid, but it represents an analytically useful framework for understanding the strengths, limitations, and trade-offs of the main families of social media experiments.

The following discussion focuses on established families of experimental archetypes. As such, we defer the formal definition and discussion of \opfes{} to Section~\ref{sec:opfes}, where they constitute the primary object of study. This organization allows us to first establish the existing experimental landscape before examining how \opfes{} reshape it.

\example{Running example}{To facilitate understanding and comparison across the established approaches, we use a common illustrative example throughout this section. We consider a hypothetical research question concerning the effect of misinformation warning labels on users’ propensity to engage with or share content. This example is chosen for its generality and relevance across multiple domains~\citep{clayton2020real,pennycookShiftingAttention2021,ecker2022psychological}. It can be implemented, with appropriate adaptations, across all the approaches we describe, illustrating how each one differently satisfies the experimental needs in practice.}

\subsubsection{Survey-Based Experiments}
\appr{Survey-based experiments} expose participants to social media content within a questionnaire environment and measure outcomes through self-reported responses such as attitudes, beliefs, or behavioral intentions \citep{blackwellHarassmentJustified2018,jhaver2019did,sharevskiSoftModeration2022,vragaNewsLiteracyMessages2022,vanDerMeerWarningsCredibility2023,hinterreiterNewsUnfold2025, chan2025cross, gligoric2019causal,cima2025contextualized, fazio2020pausing}. This category also includes feed-reconstruction designs, in which researchers embed posts or simplified feeds (e.g., screenshots or curated timelines) to approximate the structure of platform exposure. These approaches offer \prop{controllability}[high] over stimuli and randomization, enabling precise manipulation of specific features (e.g., labels, wording, ordering). At the same time, they abstract away from the interactive and social dynamics of real platforms and rely primarily on stated rather than observed behavior, which results in \prop{ecological validity}[low].

\example{Example implementation}{Participants are presented with a series of posts within an online survey, some of which include misinformation warning labels while others do not. The assignment of labeled versus unlabeled posts is randomized across participants or items. After viewing each post, participants report their likelihood of sharing it or their perceived accuracy. Outcomes are then compared across conditions to estimate the causal effect of warning labels on self-reported engagement and beliefs.}

\subsubsection{Controlled-Environment Experiments}
\appr{Controlled-environment experiments} recreate aspects of social media platforms within a researcher-designed setting, allowing participants to interact with a simulated but functional system \citep{almaliki2025combining, alvarez2018normative, celadinCivilDiscourse2024, cho2024can, ding2024counterquill, micallefFakey2021, nilsson2026study}. Unlike survey-based approaches, these environments can support actions such as browsing, clicking, liking, or sharing, thereby capturing observed behavior rather than self-reported intentions. This affords greater behavioral realism while maintaining \prop{controllability}[high] over the experimental conditions and interface. However, the environment remains artificial: participants know they are in an experiment, the social context is limited, and the mock system does not reflect the full complexity, scale, or dynamics of real platforms.

\example{Example implementation}{Participants access a web-based interface that mimics a social media feed and are asked to browse it as they normally would. The feed contains posts, some of which include misinformation warning labels while others do not, with assignment randomized at the user or post level. Participants can interact with the content by clicking, liking, or sharing within the experimental environment. The effect of labels is then estimated by comparing actual engagement behaviors across conditions.}

\subsubsection{In Situ Field Experiments on Closed Platforms}
\appr{In situ field experiments on closed platforms} are conducted directly within live social media platforms, exposing real users to experimental conditions in their natural environments. Researchers implement interventions using publicly available, user-facing platform affordances---such as posting, replying, messaging, or advertising---without access to or control over underlying system components \citep{yildirimSuspensionWarnings2023, bilewicz2021artificial, gennaro2025distribution, larsen2023counter, donati2024facebook, ajzenman2025discrimination, acquisti2020experiment}. In many cases, neither the platform nor the affected users are explicitly aware that they are part of a study, allowing researchers to observe behavior without explicit experimental framing, thus limiting risks such as observer bias or Hawthorne effect. This yields \prop{ecological validity}[high] and access to genuine behavioral responses at scale. At the same time researchers face \prop{controllability}[low], as interventions are constrained by user-facing affordances and limited control over how content is distributed. These designs may also suffer from \prop{compliance}[low] due to limitations in how ethical, legal, user consent, and transparency obligations are met.

\example{Example implementation}{Researchers create a set of controlled accounts that post content on the platform, with accounts or posts randomly assigned to include or omit misinformation warning labels. These posts are designed to resemble typical user-generated content and are published under comparable conditions (e.g., timing, topic, audience). The researchers then measure engagement with these posts, including likes, shares, comments, and reply content, to assess how users respond to the presence or absence of labels. By comparing engagement patterns across conditions, the experiment estimates the causal effect of labeling on user interaction in a real-world setting, while remaining constrained by the platform’s existing mechanisms for content distribution and visibility.}

\subsubsection{Client-Side Experiments}
\appr{Client-side experiments} modify the user's experience of a social media platform through an additional software layer that operates on top of the platform's existing application (e.g., via custom interfaces, wrappers, or extensions), without altering the platform's native client or backend systems \citep{piccardi2025reranking,bhuiyanFeedReflect2018, chang2022thread}. In practice, this layer intercepts and transforms the content delivered by the platform—such as adding labels, reordering items, or filtering posts—so that participants see a modified version of an otherwise unchanged environment. This approach enables controlled manipulation of real platform content and user interaction while preserving access to live systems. However, because the intervention is external to the platform, it is constrained by what can be modified at the interface level and cannot affect how content is generated, ranked, or distributed by the platform itself, resulting in \prop{controllability}[low]. In addition, these designs depend on the stability of the underlying application and often have \prop{operational accessibility}[low], as changes to the platform’s client or data flows can disrupt or invalidate the experimental setup. As participation requires users to adopt the additional software layer, \prop{scalability} is typically limited and subject to selection effects.

\example{Example implementation}{Participants are recruited to install a browser plugin that modifies how the social media feed is rendered in each participant's local web interface. Once installed, the plugin detects posts as they are loaded and programmatically augments them to include, or omit, misinformation warning labels, with assignment randomized at the participant or post level. The plugin operates locally in the user’s browser, without altering the platform’s underlying systems, and records engagement behaviors such as clicks, likes, shares, and comments. By comparing behavior across conditions, researchers estimate the effect of labels on user engagement in a setting that preserves real content and interaction patterns, while remaining limited to users who install the plugin and to modifications that can be applied externally to the platform’s infrastructure.}

\subsubsection{Simulation-Based Experiments}
\appr{Simulation-based experiments} study social media dynamics in fully artificial environments, where both users and interactions are modeled computationally \citep{de2019can, hata2025manipulating}. A common approach relies on ABMs, in which individual agents follow predefined behavioral rules and interact within a simulated network ~\citep{geschke2019triple, franken2021cascades, sirbu2019algorithmic, baumann2020modeling}. More recent variants incorporate generative AI, such as LLMs configured as persona-based agents (generative agent-based models, or GABMs) to produce more flexible and context-sensitive behaviors that may resemble human communication~\citep{fidone2026evaluating,tornberg2023simulating, chuang2024simulating,rossetti2026social}. Researchers may also simulate entire social networks, including content diffusion, recommendation processes, or moderation interventions. These approaches offer \prop{controllability}[high] over the environment, \prop{scalability}[high], and the ability to test counterfactual scenarios. However, their validity depends on modeling assumptions, and the absence of real users and platform dynamics limits their ability to capture actual behavior and external conditions, which results in \prop{ecological validity}[low].

\example{Example implementation}{Researchers construct a simulated social network in which agents represent users connected through a predefined or generated graph structure. Each agent produces and consumes content according to behavioral rules or generative models. Posts are randomly assigned to include or omit misinformation warning labels, and agents decide whether to engage with or share content based on their characteristics and exposure. In a generative setup, agents powered by LLMs may interpret posts and generate responses or sharing decisions in natural language. By comparing engagement and diffusion patterns across conditions, or actual responses in natural language, researchers estimate the effect of labels within the simulated system, while relying on assumptions about agent behavior and network structure.}

\subsubsection{Platform-Run Experiments}
\appr{Platform-run experiments} are conducted within social media platforms by the platform operators alone, or in collaboration with selected external researchers~\citep{gonzalez2023asymmetric,guessResharesSocialMedia2023,guessHowSocialMedia2023,nyhan2023like,allcott2024effects, kramer2014experimental}. In both cases, interventions on the live system---such as changes to ranking algorithms, recommendation systems, or moderation policies---are implemented by platform personnel. External researchers, when involved, may contribute to the design of the study and request specific modifications, but they do not have direct access to the platform’s infrastructure and cannot implement interventions independently. This arrangement enables, in principle, a wide range of system-level manipulations and large-scale observation of user behavior, offering \prop{controllability}[high], \prop{ecological validity}[high], and \prop{scalability}[high]. In practice, however, all interventions remain contingent on platform approval and priorities, thus suffering \prop{platform independence}[low]. These experiments are typically deployed without notifying users that they are part of a study, allowing observation of behavior under natural conditions, but limiting \prop{compliance}.

\example{Example implementation}{The platform silently deploys an experiment in which misinformation warning labels are added to a subset of posts within the live system, with assignment randomized at the user or post level. Because the intervention is implemented directly within the platform’s infrastructure, the labels can be seamlessly integrated into the platform's interface. The platform then measures user engagement---such as clicks, shares, and downstream diffusion---using internal data. In a partnered setting, external researchers receive the resulting experimental data from the platform and conduct the analyses on it.}

\subsection{Charting the Design Space}
\label{sec:design-space-evaluation}
The experimental archetypes previously described differ systematically in what they allow researchers to manipulate, observe, and infer. To make these differences explicit, we present a comparative assessment of the approaches in terms of the desirable properties. The resulting comparison should not be interpreted as a ranking of approaches. An approach characterized by a larger number of \protect\high~high scores is not necessarily preferable to one exhibiting more \protect\med~medium or \protect\low~low scores, as the relevance and desirability of each property depend on the specific context, research questions, and methodological priorities of a study. As such, researchers may deliberately sacrifice some properties to maximize others that are more critical for their objectives. Results of the comparative assessment are reported in Table~\ref{tab:approaches-properties}, which highlights the distinct strengths and weaknesses that characterize different approaches. For transparency, and to clarify the reasoning underlying the assigned scores, Appendix Table~\ref{tab:scores-motivations} reports a brief motivation for each score.

\begin{table*}[t]
    \centering
    \caption{Comparison of experimental approaches across desirable properties. Symbols indicate the extent to which an approach tends to support a property:~\protect\high~high,~\protect\med~medium,~\protect\low~low. A motivation for each approach--property score is reported in Appendix Table~\ref{tab:scores-motivations}.}
    \begin{minipage}[t]{0.67\textwidth}
        \centering{
        \renewcommand{\arraystretch}{1.2}
        \setlength{\extrarowheight}{1pt}
        \setlength{\tabcolsep}{0pt}
        \small
        \begin{tabular}{
            >{\raggedright\arraybackslash}p{4cm}
            *{9}{>{\centering\arraybackslash}p{0.5cm}}
        }
            
            & \multicolumn{9}{c}{\textbf{properties}} \\
            \cmidrule{2-10}
            \textbf{approaches} &
            EV &
            Ct &
            Ob &
            Cp &
            PI &
            Sc &         
            OA &
            EI &
            Rp \\
            \midrule
    		Survey-based                            & \low & \tabhigh & \med & \tabhigh & \tabhigh & \low & \tabhigh & \tabhigh & \tabhigh \\
    		Controlled-environment                 & \low	& \tabhigh & \med & \tabhigh & \tabhigh & \low & \med & \tabhigh & \tabhigh \\
    		In situ      & \tabhigh & \low & \med & \low & \med & \tabhigh & \med & \low & \med \\
    		Client-side                           & \med & \med & \med & \med & \med & \low & \low & \med & \med \\
    		Simulation-based                      & \low & \tabhigh & \tabhigh & \tabhigh & \tabhigh & \tabhigh & \low & \tabhigh & \tabhigh \\
    		Platform-run                           & \tabhigh & \tabhigh & \tabhigh & \med & \low & \tabhigh & \tabhigh & \low & \low \\
            \rowcolor{opfegrey}\textbf{OPFE}  & \tabhigh & \med & \tabhigh & \tabhigh & \med & \med  & \low & \low & \med \\
            \bottomrule
        \end{tabular}
        }
    \end{minipage}
    \hfill
    \begin{minipage}[t]{0.30\textwidth}
    \footnotesize
    \centering{
        \renewcommand{\arraystretch}{1.3}
        \setlength{\tabcolsep}{4pt}
        \begin{tabular}{@{}ll@{}}
            
            EV & Ecological Validity \\
            Ct & Controllability \\
            Ob & Observability \\
            Cp & Compliance \\
            PI & Platform Independence \\
            Sc & Scalability \\
            OA & Operational Accessibility \\
            EI & Experimental Isolation \\
            Rp & Reproducibility \\
            
        \end{tabular}
    }
    \end{minipage}

    \label{tab:approaches-properties}
\end{table*}

\subsubsection{Comparative Analysis of the Experimental Approaches}
\label{sec:design-space-evaluation-comparative-analysis}
Several patterns emerge from the comparisons in Table~\ref{tab:approaches-properties}. First, approaches such as \appr{survey-based}, \appr{controlled-environment}, and \appr{simulation-based experiments} tend to maximize \prop{controllability}, \prop{reproducibility}, \prop{experimental isolation}, and \prop{compliance}, at the expense of \prop{ecological validity}. These approaches enable precise manipulation and measurement under relatively stable conditions, but they rely on simplified representations of social media environments, user behavior, or network dynamics. As a result, they are well suited for studying tightly controlled causal mechanisms, but less effective at capturing the complexity of real-world platform interactions. Interestingly, despite occupying similar regions of the design space, these approaches differ substantially in \prop{operational accessibility}. \appr{Survey-based experiments} are comparatively lightweight and broadly accessible, whereas large-scale simulation frameworks---particularly those based on generative AI agents or synthetic social networks---require sophisticated computational infrastructures. This difference has shaped disciplinary adoption patterns, with simulations being more prevalent in computer science and survey-based approaches being predominantly used in less technically intensive communities such as communication and psychology~\citep{fan2023new}.

Approaches conducted directly within live platforms, including \appr{in situ field experiments on closed platforms} and \appr{platform-run experiments}, occupy the opposite region of the space. These approaches provide \prop{ecological validity}[high] and \prop{scalability}[high], as they operate on real infrastructures and populations. However, they also exhibit structural limitations in terms of \prop{experimental isolation}[low] and \proph{\protect\med~medium} or \prop{platform independence}[low], \prop{reproducibility}, and \prop{compliance}. In particular, \appr{platform-run experiments} concentrate experimental capabilities within platforms themselves, while independent \appr{in situ field experiments} are constrained by publicly accessible affordances, lower \prop{observability}, and possible tensions with platform policies or ethical requirements, resulting in \prop{compliance}[low]. A third category consists of \appr{client-side experiments}, which occupy an intermediate position across most dimensions. Rather than maximizing any single property, this approach attempts to partially reconcile realism and control through an adaptive experimental setup. This flexibility can make it practically useful in a variety of contexts, but it also means that it inherits limitations from both controlled and on-platform methodologies.

\begin{figure}[t]
    \centering
    \includegraphics[width=1\textwidth]{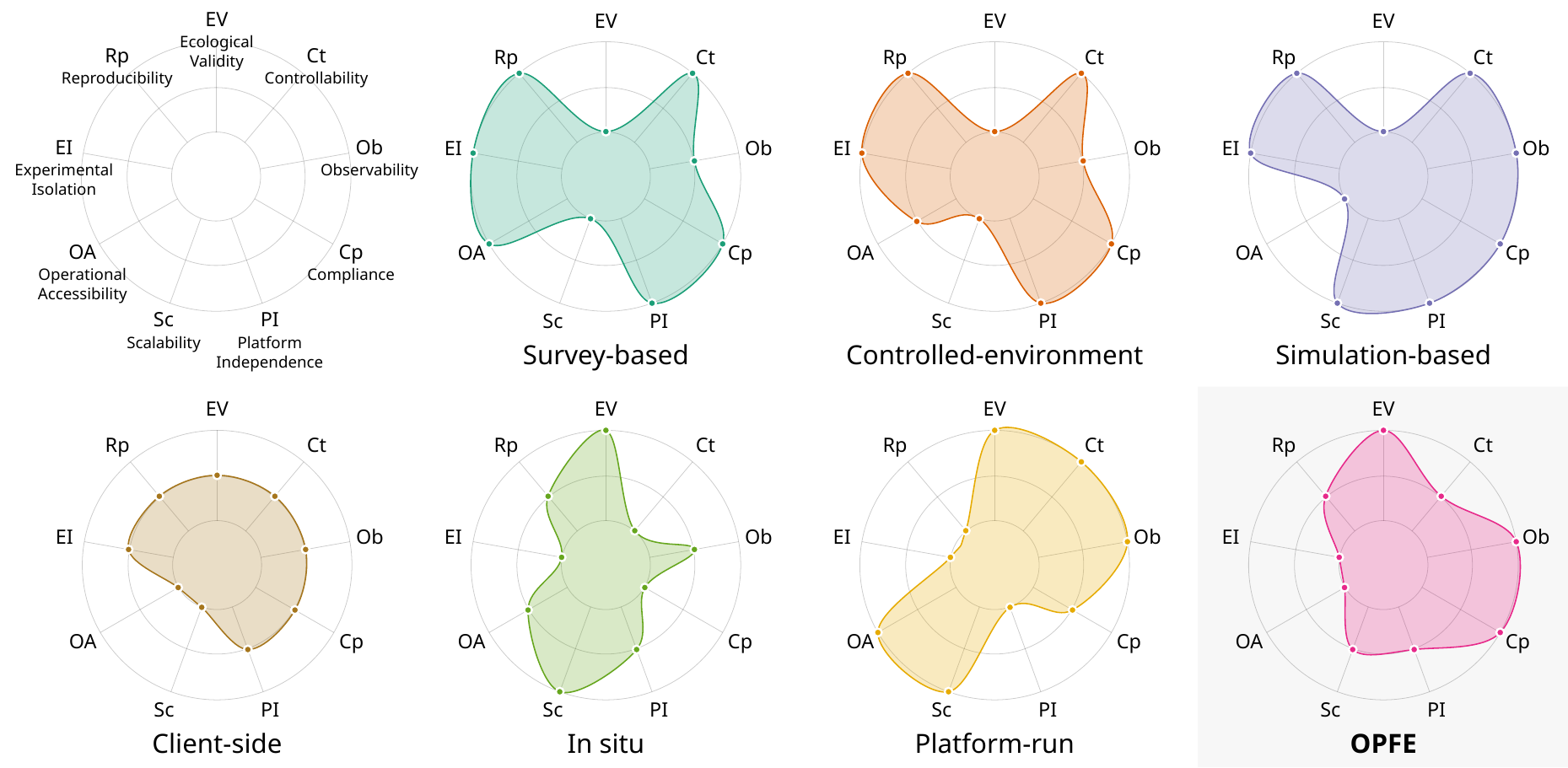}
    \caption{Methodological profiles of the considered experimental archetypes. Each radar chart visualizes the relative support provided by an experimental archetype across the desirable properties. Unlike Table~\ref{tab:approaches-properties}, which reports the individual property scores, these profiles highlight the overall balance of strengths and limitations that characterizes each approach, making their distinctive methodological trade-offs apparent. The archetypes are listed in ascending order of \prop{ecological validity} (EV) and decreasing order of \prop{platform independence} (PI).}
    \Description{A figure depicting eight radar charts: one for each experimental archetype and another serving as a legend to explain the order of the properties. The values shown in the radar chart of each experimental archetype are the same as those presented in Table 2.}
    \label{fig:radar}
\end{figure}

The comparisons reported in Table~\ref{tab:approaches-properties} and the methodological profiles visualized in Figure~\ref{fig:radar} and Figure~\ref{fig:nmds_design_space} highlight how different approaches occupy distinct regions of the design space and embody characteristic trade-offs. These differences explain why social media experimentation has historically evolved through a fragmented ecosystem of partially complementary methodologies. Furthermore, these patterns also reveal a broader structural asymmetry in the current landscape of social media experimentation. Approaches that maximize realism often do so by sacrificing \prop{reproducibility}, \prop{platform independence}, or \prop{compliance}, while approaches that maximize \prop{controllability} and \prop{reproducibility} typically abstract away from the very platform dynamics they seek to study. The resulting design space is therefore characterized by the absence of approaches capable of simultaneously combining \prop{ecological validity}[high] with\protect\med~medium or\protect\high~high support for \prop{platform independence}, \prop{compliance}, \prop{reproducibility}, and \prop{controllability}.

Beyond the comparative analysis, this framework can also inform the design of future experiments. Researchers may use it to identify the experimental approach that best satisfies the methodological needs and priorities of a particular study, or alternatively, to reason about the kinds of research questions and experimental designs that are enabled by a given approach. In this sense, the framework serves not only as an analytical tool for comparing existing methodologies, but also as a practical guide for selecting and designing future social media experiments.

\begin{figure}[t]
    \centering
    \begin{minipage}[c]{0.5\textwidth}
        \centering
        \includegraphics[width=\linewidth]{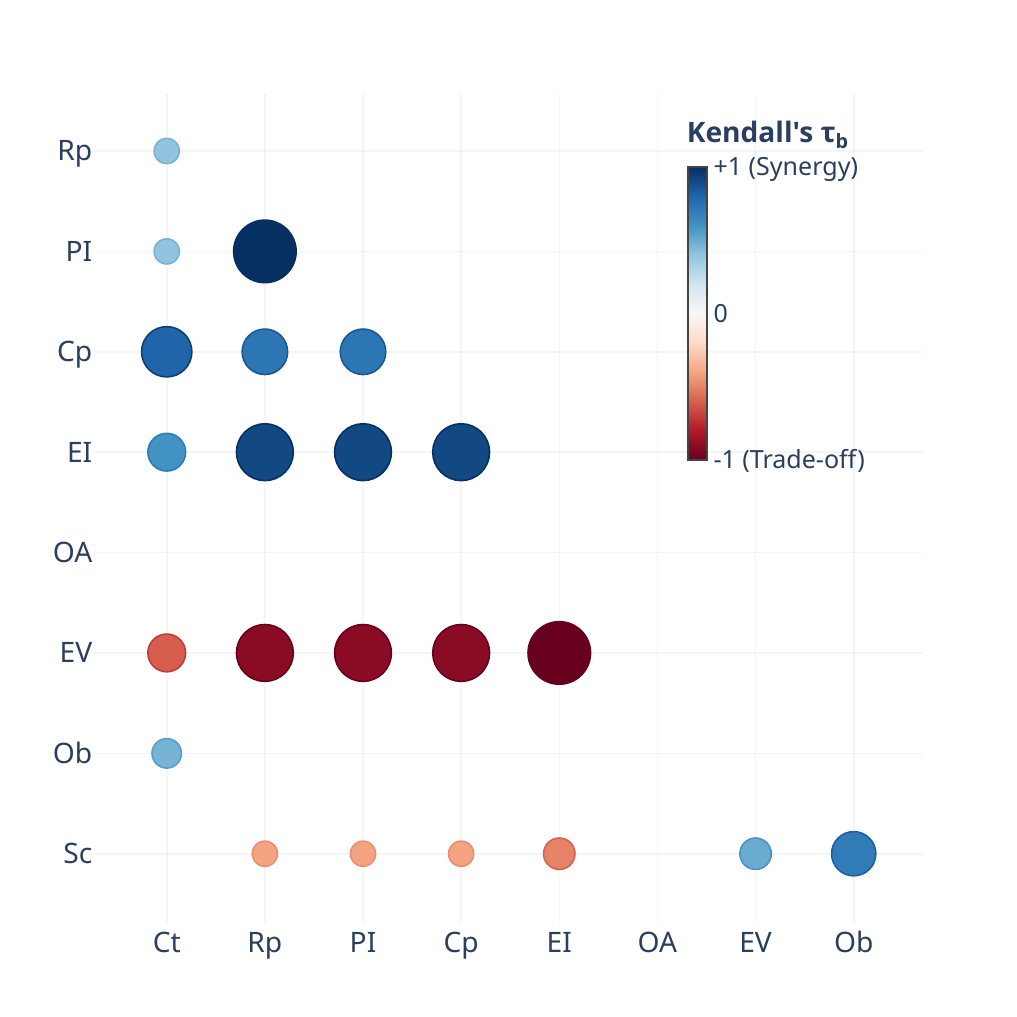}
    \end{minipage}\hspace{2em}
    \begin{minipage}[c]{0.28\textwidth}
        \footnotesize
        \centering
        \renewcommand{\arraystretch}{1.3}
        \setlength{\tabcolsep}{4pt}
        \begin{tabular}{@{}ll@{}}
            EV & Ecological Validity \\
            Ct & Controllability \\
            Ob & Observability \\
            Cp & Compliance \\
            PI & Platform Independence \\
            Sc & Scalability \\
            OA & Operational Accessibility \\
            EI & Experimental Isolation \\
            Rp & Reproducibility \\
        \end{tabular}
    \end{minipage}
    
    \caption{Correlations among desirable properties, highlighting methodological synergies and trade-offs. Correlations are computed using Kendall's $\tau_b$ rank correlation coefficient over the ordinal comparative assessments reported in Table~\ref{tab:approaches-properties}. Blue bubbles indicate positive correlations (synergies), where properties tend to increase or decrease together, whereas red bubbles indicate negative correlations (trade-offs), where improving one property is typically associated with reducing another. Bubble size and color intensity are proportional to the absolute value of the correlation. To emphasize the most salient relationships, only correlations with $|\tau_b| \geq 0.40$ are shown. To characterize the strengths and limitations of the existing methodological landscape, the analysis excludes \opfes{}.}
    \Description{A correlation heatmap showing the relationship between properties across experimental archetypes, excluding OPFEs. Both the x-axis and y-axis list the properties, and each cell displays their correlation, calculated by converting High, Medium, and Low scores into numerical values. Synergies (positive correlations) and trade-offs (negative correlations) are discussed in detail in the text.}
    \label{fig:properties-tensions}
\end{figure}

\subsubsection{Interdependencies and Trade-Offs of the Desirable Properties}
\label{sec:design-space-evaluation-trade-offs}
Beyond enabling the comparison of experimental approaches, Figure~\ref{fig:radar} also provides a useful lens for examining the relationships among the desirable properties themselves. In particular, it highlights several recurring interdependencies and trade-offs that shape the design space of social media experiments. For instance, while the properties introduced in Section~\ref{sec:design-space-properties} are defined independently for analytical clarity, they are not fully independent in practice. Experimental approaches are inherently shaped by trade-offs where improving one dimension comes at the expense of another. Figure~\ref{fig:properties-tensions} illustrates a subset of common synergies and tensions between properties, highlighting how design choices may simultaneously enable and constrain different aspects of experimental research. These relationships are not universal, but reflect recurring patterns that emerge across the considered experimental archetypes.

As shown in Figure~\ref{fig:properties-tensions}, several tensions and synergies are particularly salient and cluster around a small set of central properties. The figure reveals a prominent block of synergistic properties, indicating that experimental approaches tend to rank similarly across these dimensions. This block comprises \prop{experimental isolation}, \prop{reproducibility}, \prop{platform independence}, \prop{compliance}, and \prop{controllability}. Within it, \prop{reproducibility} and \prop{platform independence} exhibit perfect alignment, serving as the primary anchors of the cluster. These synergies are not coincidental but instead reflect fundamental characteristics of experimental design. In the absence of explicit collaboration or structural openness from a platform, achieving high \prop{controllability} and \prop{experimental isolation} typically requires researchers to conduct experiments in abstracted environments. By contrast, experiments conducted directly on real platforms offer only limited opportunities to control experimental conditions or observe the variables of interest. Moreover, because abstracted environments generally require active participant recruitment, they naturally support informed consent and ethical oversight, resulting in high \prop{compliance}. \prop{Operational accessibility} also aligns with this block, although more weakly, likely because some highly controlled approaches, such as surveys, are broadly accessible, whereas others, such as simulations, require substantial technical expertise and infrastructure.

Crucially, maximizing this block of properties comes at a substantial cost to \prop{ecological validity}. This reflects a fundamental structural trade-off: the abstractions required to achieve isolation, independence, and compliance necessarily remove many of the complex dynamics that characterize real social media platforms and human behavior. Conversely, increasing \prop{ecological validity} requires experimentation in real-world settings, where interactions among users, algorithms, and platform policies inevitably limit \prop{experimental isolation}, \prop{controllability}, \prop{reproducibility}, and \prop{platform independence}. As a result, although \prop{ecological validity} is among the most desirable properties for scientific relevance, no existing experimental approach can achieve it without compromising many of the other desirable properties.
Taken together, these relationships demonstrate that no single experimental archetype can simultaneously maximize all desirable properties. Instead, experimental design is fundamentally an exercise in navigating unavoidable trade-offs across multiple methodological dimensions.

The relationships discussed above describe structural interactions among the properties. A complementary consideration concerns how these properties should be interpreted when selecting or evaluating an experimental approach. In fact, while the set of properties we introduced is fixed, their relative importance is inherently context-dependent. Different research communities, application domains, and regulatory environments may prioritize properties differently. For example, \prop{compliance} may be particularly salient for researchers operating under stringent regulatory frameworks, such as in the European Union, where the evolving legal landscape can both hinder and help academic research~\citep{leerssen2023scraping}. In other jurisdictions, lighter constraints on data access and privacy~\citep{lubin2024mapping} can allow researchers to place greater emphasis on \prop{scalability} or \prop{operational accessibility}. Similarly, fields such as machine learning may prioritize \prop{scalability}, whereas computational social science may place greater weight on \prop{ecological validity}. The proposed set of properties thus provides a flexible basis for assessing the strengths and limitations of experimental approaches in light of specific research questions, methodological priorities, and institutional constraints.

The context-dependent interpretation of the properties also motivates a brief reflection on the scope of the proposed framework. The considered set of properties is not intended to exhaustively characterize every aspect of social media experimentation. Rather, it captures a small set of methodological dimensions that most directly shape researchers' ability to generate causal evidence across different classes of experimental approaches. Accordingly, the framework emphasizes properties that are intrinsic to broad methodological families rather than to individual experimental instances. For example, the duration of a study is not treated as a standalone property because it primarily reflects implementation choices rather than the capabilities of the underlying approach. Likewise, although the proposed properties are defined to capture distinct aspects of the experimental design space, some factors naturally influence multiple dimensions simultaneously. The temporal stability of platforms, user populations, or data interfaces, for instance, affects \prop{reproducibility}, \prop{observability}, and \prop{operational accessibility} at the same time. However, rather than introducing such cross-cutting factors as additional properties, the framework captures their influence through the methodological dimensions they affect. This abstraction favors analytical clarity and parsimony while preserving the ability to reason about the multifaceted constraints that shape social media experimentation.

\section{Open Platform Field Experiments}
\label{sec:opfes}

The emergence of open social media has introduced new possibilities for conducting experiments directly within real-world online environments. The increasing availability of open protocols and externally accessible platform data and components enables forms of experimentation that are difficult or outright impossible to conduct on closed platforms. We refer to these approaches as \textit{open-platform field experiments} (\opfes).

\opfes are conducted within social media platforms that expose meaningful platform components---such as protocols, clients, data, ranking or moderation systems---in ways that allow external actors to access and modify them independently. Researchers must therefore not only implement the experimental intervention, but also make the resulting components available to participants and sustain the infrastructure through which the experiment is conducted. However, unlike \appr{platform-run experiments}, \opfes do not require discretionary collaboration or approval from platform providers. Unlike \appr{in situ field experiments on closed platforms}, they are not limited to publicly available features used as-is, but enable direct intervention on the open parts of the platform infrastructure. At the same time, \opfes also differ from \appr{client-side experiments}, where researchers build overlays or extensions on top of existing applications, without modifying the underlying system. Instead, they operate on functional components of live social media platforms, allowing experiments with actual users and interactions, while retaining a substantial degree of autonomy and transparency. This autonomy, however, is bounded. Researchers can act only on the open components of an evolving technical architecture, which introduces new forms of technical and methodological complexity that we examine below.

\subsection{OPFEs in the Experimental Design Space}
\label{sec:opfes-design-space}
\opfes represent a distinct class of experimental approaches that combines \prop{ecological validity}[high] with strong support for \prop{platform independence}, \prop{controllability}, and \prop{observability}. The bottom row of Table~\ref{tab:approaches-properties} positions \opfes within the experimental design space, enabling direct comparison with the other archetypes of social media experiments. As shown, only three classes of approaches achieve \prop{ecological validity}[high]: \textit{(i)} \appr{platform-run experiments}, \textit{(ii)} \appr{in situ field experiments on closed platforms}, and \textit{(iii)} \opfes. These approaches support \prop{ecological validity} through different trade-offs. \appr{Platform-run experiments} benefit from direct access to the underlying infrastructure and user population, but exhibit \prop{platform independence}[low] and \prop{reproducibility}[low] due to their reliance on discretionary platform collaboration and proprietary systems. \appr{In situ field experiments on closed platforms} retain greater researcher autonomy (i.e., \prop{platform independence}[medium]), but are constrained by \prop{controllability}[low], possible tensions with ethical and legal requirements that often result in \prop{compliance}[low], and limited \prop{observability}. \opfes rebalance these trade-offs. By enabling direct intervention on open components of live social media platforms, they make it possible to conduct ecologically valid experiments while simultaneously supporting stronger forms of \prop{controllability}, \prop{observability}, \prop{platform independence}, and \prop{compliance}. This set of features has historically been difficult to achieve in social media experiments, which positions \opfes in an unoccupied region of the design space.

\begin{figure}[t]
    \centering
    \includegraphics[width=0.65\textwidth]{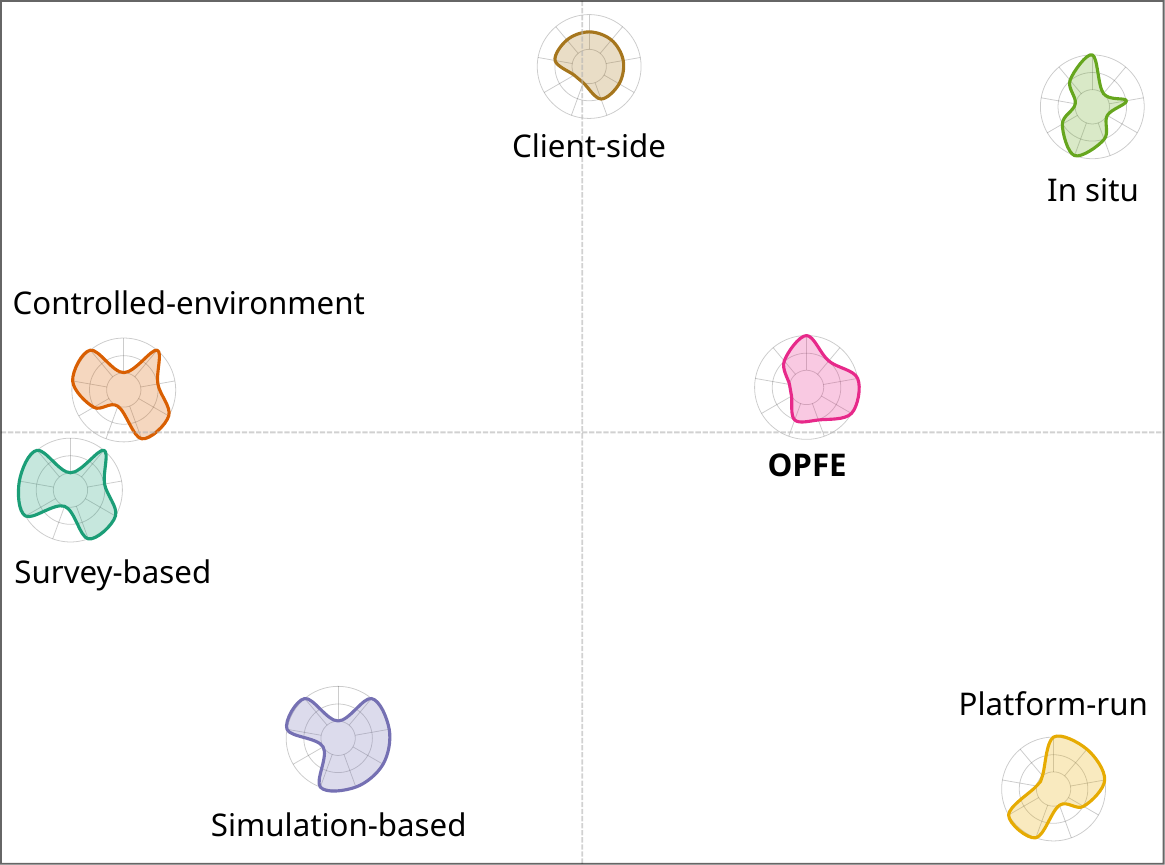}
    \caption{Two-dimensional projection of the experimental design space induced by the scores in Table~\ref{tab:approaches-properties}. As shown, \opfes colonize an unoccupied region of the design space, bridging capabilities that have traditionally been separated across existing methodologies. The projection is obtained by mapping the ordinal property assessments (\protect\low~low, \protect\med~medium, \protect\high~high) to numerical values, computing pairwise Manhattan distances between approaches, and embedding the resulting dissimilarity matrix into two dimensions using non-metric multidimensional scaling (NMDS).
    The spatial arrangement reflects the overall similarity of approaches with respect to the desirable properties, rather than absolute quantitative distances.}
    \Description{A dimensionality reduction projection of the presented experimental archetypes in 2D, showing seven points, one for each archetype. Each point is represented by the corresponding radar chart from Figure 1. The spatial distribution of this projection is discussed in detail in the text.}
    \label{fig:nmds_design_space}
\end{figure}

Figure~\ref{fig:nmds_design_space} provides a visual summary of the comparative analysis presented in Table~\ref{tab:approaches-properties} and Figure~\ref{fig:radar}. Existing experimental approaches naturally organize into two broad regions of the design space. On one side lie highly controlled methodologies, such as \appr{survey-based}, \appr{controlled-environment}, and \appr{simulation-based experiments}, which emphasize \prop{controllability}, \prop{reproducibility}, and \prop{experimental isolation}, but sacrifice \prop{ecological validity}. On the other side are approaches conducted directly within live platforms, including \appr{platform-run experiments} and \appr{in situ field experiments on closed platforms}, which maximize \prop{ecological validity} while accepting limitations in \prop{platform independence}, \prop{reproducibility}, \prop{compliance}, or \prop{controllability}. \appr{Client-side experiments} occupy an intermediate position between these two extremes. Notably, \opfes do not simply extend either family nor interpolate between them. Instead, by rebalancing several of the methodological trade-offs identified in Section~\ref{sec:design-space-evaluation}, they colonize a previously unoccupied region of the design space, illustrating their role as a distinct experimental paradigm. At the same time, \opfes do not completely eliminate the existing tensions, but rather shift them toward new directions. Similar to other live platform experiments, \opfes exhibit \prop{experimental isolation}[low], as experiments unfold within shared social and algorithmic environments where spillovers and interdependencies are inevitable~\citep{eckles2017design,pedreschi2025human}. Their \prop{controllability} stops at the boundary of the open components, hence the \protect\med~medium score, since researchers cannot intervene on parts of the platform that remain closed. \opfes also feature \prop{platform independence}[medium], as researchers must come to terms with the underlying architecture and the applicable terms of use.  Similarly, \prop{scalability} can be capped by the need for active opt-in: because researchers typically lack the capacity to silently enroll users, participants usually have to explicitly consent and connect to the experimental environment provided by the study.
Moreover, the support of \opfes for \prop{reproducibility} extends only to the open technical apparatus, as the users, interactions, and live platform dynamics cannot be accurately recreated over time. These trade-offs are finally compounded by \prop{operational accessibility}[low], as designing and deploying \opfes may require substantial technical expertise and familiarity with possibly complex and evolving platform infrastructures~\citep{kleppmannBlueskyProtocolUsable2024}. \opfes thus democratize \textit{institutional} access, but not necessarily \textit{technical} access. Nonetheless, the set of features that characterize \opfes opens up the possibility to study causal questions that were historically confined either to controlled but artificial settings or to opaque platform-mediated experimentation.

\example{Example implementation}{Having established the notion of OPFEs, we now instantiate the running example introduced in Section~\ref{sec:design-space-approaches} within this experimental paradigm. Researchers implement a modified feed or client component that adds misinformation warning labels to a subset of posts, with assignment randomized at the user or post level. Users accessing the platform through the modified component encounter labeled and unlabeled content as part of their normal experience, while engagement variables---such as likes, shares, comments, or reposts---are measured natively through the available data interfaces. This implementation combines capabilities of both live platform experiments---since it operates within a real social media environment with actual users and native behavioral measurements, and controlled experimental settings---by allowing direct intervention on open platform components.}

\subsection{Open Platform Elements}
\label{sec:opfes-layers}
\opfes occupy a distinctive region of the experimental design space by combining methodological properties that are otherwise difficult to achieve simultaneously. This distinctive positioning stems from the functional characteristics of open platforms. In turn, these characteristics emerge from different functional aspects of the platform, which can be understood in terms of a small set of platform elements, including users and the social environment, data interfaces, modifiable components, and platform policies. As shown in Table~\ref{tab:elements-properties}, each element exposes specific capabilities relevant to experimentation, whose extent depends on characteristics such as openness, interoperability, transparency, and persistent availability. Collectively, these capabilities determine which experimental needs can be satisfied and, consequently, why \opfes naturally exhibit a distinctive combination of desirable properties.

\begin{table*}[t]
    \centering
    \caption{Contributions of the open platform and the relevant external elements to the desirable properties of \opfes{}. Green upward arrows indicate properties that an element tends to facilitate, whereas red downward arrows indicate properties that it tends to hinder. For each element, the properties most strongly associated with it are highlighted in bold. The desirable properties of \opfes{} emerge from the combined contribution and interaction of multiple elements rather than from any single element in isolation. The table therefore provides an architectural interpretation of how the technical and governance characteristics of open platforms give rise to the overall methodological profile of \opfes{} reported in the last table row.}
    \begin{minipage}[t]{0.67\textwidth}
        \centering{
        \renewcommand{\arraystretch}{1.2}
        \setlength{\extrarowheight}{1pt}
        \setlength{\tabcolsep}{0pt}
        \small
        \begin{tabular}{
            >{\centering\arraybackslash}m{0.75cm}
            >{\raggedright\arraybackslash}p{4cm}
            *{9}{>{\centering\arraybackslash}p{0.5cm}}
        }
            && \multicolumn{9}{c}{\textbf{properties}} \\
            \cmidrule{3-11}
            \multicolumn{2}{c}{\textbf{elements}} &
            EV &
            Ct &
            Ob &
            Cp &
            PI &
            Sc &         
            OA &
            EI &
            Rp \\
            \midrule
            \multirow{2}{*}{\rotatebox[origin=c]{90}{\textit{external}}}
            & crowdsourcing \& ads          & \propdown &  &  &  &  & \propupB &  &  &  \\
    		& applicable laws               &  &  &  & \propupB &  &  &  &  &  \\
            \midrule
            \multirow{4}{*}{\rotatebox[origin=c]{90}{\textit{open platforms}}}
    		& users \& social environment   & \propupB &  &  &  &  & \propupB &  & \propdown & \propdown \\
    		& data                          &  &  & \propupB & \propdown &  &  & \propdown &  & \propup \\
    		& modifiable components         & \propup & \propupB & \propup & \propup & \propdown & \propdown & \propdown &  & \propup \\
    		& policy \& governance          &  &  &  & \propupB & \propdown &  &  &  & \propup\ \propdown \\
            \midrule
            \rowcolor{opfegrey}& \textbf{OPFE} & \tabhigh & \med & \tabhigh & \tabhigh & \med & \med  & \low & \low & \med \\
            \bottomrule
        \end{tabular}
        }
    \end{minipage}
    \hfill
    \begin{minipage}[t]{0.30\textwidth}
    \footnotesize
    \centering{
        \renewcommand{\arraystretch}{1.3}
        \setlength{\tabcolsep}{4pt}
        \begin{tabular}{@{}ll@{}}
            EV & Ecological Validity \\
            Ct & Controllability \\
            Ob & Observability \\
            Cp & Compliance \\
            PI & Platform Independence \\
            Sc & Scalability \\
            OA & Operational Accessibility \\
            EI & Experimental Isolation \\
            Rp & Reproducibility \\
        \end{tabular}
    }
    \end{minipage}
    \label{tab:elements-properties}
\end{table*}

\subsubsection{Users and Social Environment}
Unlike approaches that require recruiting participants into artificial settings, \opfes can leverage an existing population of users who naturally inhabit the platform. The users and social environment element comprises the platform's user population together with the social context in which interactions occur, including active and inactive users, communities, organic behavioral dynamics, as well as the social (i.e., follower--followee relationships) and interaction (e.g., likes, comments, and reshares) networks. Note that the availability of a large existing user base does not prevent researchers from recruiting participants through alternative channels, such as crowdsourcing or on- and off-platform advertising~\citep{bradyRedesigningAlgorithms2026}.

By inheriting an existing user base, \opfes enable both organic recruitment and naturalistic exposure, allowing participants to encounter experimental interventions as part of their ordinary platform experience. This reduces the need to assemble a study population from scratch and mitigates issues such as participant dropout, unfamiliarity with the experimental environment, or limited engagement, thereby facilitating longitudinal participation. Because participants remain embedded in their existing social networks, this element also enables experiments that investigate social contagion, spillover effects, and other forms of network-mediated influence~\citep{eckles2017design}. Furthermore, when combined with open data, the existing user population enables large-scale observation of real behavioral processes as they unfold in situ. This positions \opfes between those experimental approaches that require researchers to recruit their own participants, and those conducted on large closed platforms, which benefit from substantially larger existing user populations. Consequently, this element supports \prop{ecological validity} and \prop{scalability}, although the available participant pool remains bounded by the size and adoption of the underlying open platform. At the same time, the presence of persistent social ties and interaction networks reduces \prop{experimental isolation}, as interventions may propagate beyond the directly treated participants, and limits \prop{reproducibility}, since the same social environment may not be faithfully recreated across repeated studies.

\subsubsection{Data}
The data element comprises the platform data itself---including raw content, behavioral traces, and metadata---together with the set of interfaces and components that allow anyone to independently access them. The latter may include application programming interfaces (APIs), public data streams (e.g., firehose services), and data repositories. Although the specific technologies vary across platforms, the defining characteristic of this element is that it exposes platform data through stable, documented, and externally accessible interfaces. However, openness should not be interpreted as \textit{unrestricted} access. Open platform providers still determine which data are exposed and under what conditions. Rather, open platforms typically impose substantially fewer constraints than their closed counterparts, enabling broad and independent access through transparent and reusable mechanisms. As such, access is generally subject to few restrictions on the types and volume of data that can be collected, temporal coverage, query capabilities, rate limits, monetary fees, and eligibility requirements, with little or no reliance on ad hoc agreements and discretionary platform approval.

By exposing behavioral traces generated during platform use, it enables large-scale behavioral measurement, longitudinal observation of users and communities, and the reconstruction of interaction dynamics. In combination with the users and social environment element, it also supports the observation of naturally occurring behavior at scale. Furthermore, the availability of accessible datasets and standardized data interfaces facilitates \prop{reproducibility}, while enabling independent auditing and validation of platform processes such as content moderation, information diffusion, or algorithmic ranking. Therefore, beyond their methodological value, these capabilities also contribute to the transparency and accountability of online platforms~\citep{kaushal2024automated,trujillo2025dsa}. At the same time, the data element introduces new technical and regulatory challenges. Effectively collecting, integrating, and managing large-scale platform data may require substantial technical expertise, thereby reducing \prop{operational accessibility}. Moreover, access to rich behavioral data raises important privacy, ethical, and legal considerations that must be carefully addressed throughout the experimental lifecycle, placing additional demands on \prop{compliance}. Overall, this element is the primary driver of the high \prop{observability} afforded by \opfes while simultaneously introducing many of the practical challenges associated with conducting experiments on open social media platforms.

\subsubsection{Modifiable Components}
The modifiable components comprise the functional parts of an open platform on which researchers can directly intervene to realize experimental manipulations. This element thus constitutes the experimental intervention surface of an \opfe. Its defining characteristic is that the implementation of selected platform components is openly available for independent inspection and reuse, for example, through public software repositories. Researchers can therefore create modified versions of existing platform components, either to implement new intervention mechanisms or to deploy experimental treatments through existing platform functionalities. These modifications do not alter the platform's default behavior for all users. Rather, they create alternative implementations that coexist with the original components, allowing researchers to experimentally compare the default and modified versions under controlled conditions. This capability naturally supports the construction of experimental counterfactuals, as participants interacting with the default component experience the ordinary platform behavior while those exposed to a modified version experience an experimental condition.

Because modified components coexist with their default counterparts, researchers must ensure that some participants interact with them. In fact, creating a modified component does not replace the corresponding platform functionality. Unless users explicitly choose to switch to the modified implementation, they continue interacting with the default one and therefore experience the ordinary platform behavior. As a result, researchers must either rely on the spontaneous adoption of the modified component by existing platform users, or actively encourage selected participants to use it. In the former case, researchers simply make the modified component available and wait for spontaneous adoption. In the latter, they may actively encourage adoption, for example to accelerate recruitment or target specific user populations. In this case, participant recruitment becomes more similar to those experimental approaches that mandate recruitment, although researchers may still benefit from recruiting from native platform users rather than assembling an entirely different participant base. This flexibility allows \opfes to balance researcher control against the advantages of organic participation. The explicit adoption of modified components also creates a natural opportunity to satisfy ethical and legal requirements. Whether users discover the modified implementation organically or are directly invited to use it, the adoption process provides a convenient point at which researchers can present study information, obtain informed consent, and provide other participant-facing documentation. In this sense, the same mechanism that limits spontaneous exposure also facilitates transparent participation, contributing to the comparatively strong \prop{compliance} of \opfes. At the same time, because implementing and maintaining modified platform components requires intervening on production-quality software and evolving platform infrastructures, this element strongly supports \prop{controllability} while reducing \prop{operational accessibility}, as deploying such interventions may demand substantial technical expertise. Finally, the availability of modified components in the form of code repositories implementing the modifications enables the sharing of experimental implementations strengthening the \prop{reproducibility} of \opfes.

The specific methodological opportunities enabled by this element depend on which platform components are modified, three key examples of which are discussed below.

\paragraph{Client and Interface.} Client and interface components constitute the primary interaction surface between users and the platform. They include clients (e.g., app, web, mobile), user interfaces, notification systems, and more generally, all mechanisms through which content and platform functionalities are presented to the users. This component primarily supports interface interventions \citep{seering2019designing}. However, since it is directly responsible for how users experience the platform, it also enables precise control over content exposure and randomized presentation conditions. Even when the interface itself is not the object of study, modifying it may allow researchers to determine which users are exposed to treatment, when they are exposed, and how interventions are presented. Finally, because every user interaction passes through this component, it substantially extends the measurement capabilities provided by the data element. Beyond recording observable platform actions, modified clients can collect fine-grained client telemetry, capturing ephemeral, volatile, and otherwise inaccessible behavioral traces, including passive content views, scrolling behavior, reading time, cursor movements, draft composition and abandonment, edit histories, and other transient user behaviors that are typically unavailable through conventional platform data interfaces. This richer observability supports more detailed analyses of user behavior while preserving experimentation within an ecologically valid environment.

\paragraph{Feed and Recommendation.} Feed and recommendation components determine which content is presented to users and in what order. They include feed-generation systems, ranking and recommendation algorithms, content selection mechanisms, personalization services, and custom feed generators. These components are central to contemporary social media, as they shape users' information environments and have become the subject of growing scientific, societal, and regulatory attention due to their influence on information exposure, engagement, and public discourse \citep{twenge2018increases,guessHowSocialMedia2023,bradyRedesigningAlgorithms2026}. Modifying these components enables researchers to intervene directly on users' information environments. This supports manipulation of content exposure, randomized ranking interventions, algorithmic experiments, and the study of recommender system effects. It also enables investigations of amplification, visibility, and personalization dynamics by experimentally varying which content is selected, prioritized, or recommended to different users. Compared with client-side interventions, modifications to feed and recommendation components act on the mechanisms that determine information exposure itself, rather than on how that information is presented. Consequently, they provide a particularly powerful means of studying the causal effects of algorithmic curation.

\paragraph{Labeling and Moderation.} Labeling and moderation components govern the enforcement of platform rules by determining when the ordinary visibility of content, user interactions, or accounts should be modified in response to violations of community guidelines, platform policies, or applicable laws. They include automated or human-assisted moderation systems, content labeling services, filtering mechanisms, visibility restrictions, warning messages, informational labels, and other mechanisms that alter how users interact with content or with one another. These interventions can be applied to individual posts, accounts, and communities, or platform-wise. Unlike feed and recommendation components, which determine what content is promoted or prioritized, moderation components determine what content or behaviors are permitted, restricted, or accompanied by contextual information. These components have become a central focus of scientific research, public debate, and regulatory initiatives, reflecting growing concerns over the governance of online platforms and the societal consequences of moderation decisions \citep{alvarez2018normative,sharevskiSoftModeration2022,guessResharesSocialMedia2023}. Their modification enables researchers to experimentally evaluate moderation interventions under real-world conditions, and more broadly, they support the study of how different strategies influence user behavior, content dissemination, compliance with platform rules, and the overall dynamics of online communities. By making these mechanisms directly modifiable, open platforms enable independent and reproducible evaluations of moderation systems that have historically been difficult to conduct on proprietary social media platforms.

\subsubsection{Policy and Governance}
The policy and governance element comprises the institutional framework that regulates how the platform, its data, and its technical infrastructure may be used. It includes platform-wide documents such as terms of service, developer and API policies, data-access policies, technical documentation, and possible research or academic agreements, as well as community-specific rules that may govern particular spaces within the platform. For platforms exposing modifiable components, this element may also specify which components can be modified, how modifications should be implemented, and the conditions under which they may be deployed.

Unlike the previous elements, which describe the platform's technical capabilities, this element defines the institutional conditions under which those capabilities can be exercised. Access to the platform, its data, or its development infrastructure is typically conditional upon accepting these rules, which are established unilaterally by the platform provider and may evolve over time. Consequently, they determine the scope of permissible experimentation and introduce a degree of dependence on platform governance. Failure to comply may result in restrictions or revocation of access to platform resources, and in some circumstances may also have legal consequences~\citep{ogrady2025unethical}.

By clearly specifying the rights, obligations, and operational boundaries applicable to researchers, this element provides the basis for conducting transparent and compliant experiments. However, while open platforms provide an infrastructure that is more compatible with satisfying these obligations, they do not remove them. Compliance with platform policies alone is generally not sufficient to ensure the overall compliance of an experiment, as this element represents only one block of the broader ethical and legal framework governing research and online platforms~\citep{lubin2024mapping}. Researchers must therefore also comply with applicable regulations, institutional requirements, and disciplinary ethical standards, such as data protection legislation (e.g., the EU's GDPR), institutional review board (IRB) approval, and research ethics protocols (e.g., informed consent, opt-in and opt-out mechanisms). At the same time, because these policies and their associated technical documentation explicitly describe the platform governance, they also facilitate \prop{reproducibility} by enabling other researchers to understand and reconstruct the experimental environment. Conversely, evolving policies, changing documentation, or modifications to platform governance may limit \prop{reproducibility} over time and constrain the range of experiments that remain feasible~\citep{bruns2021after,freelon2018computational}.

\section{OPFEs on Bluesky}
\label{sec:bluesky}
The previous section characterized \opfes independently of any specific platform. We now instantiate this conceptual model on \bsky{} and the \atproto{}, illustrating how the general elements and capabilities discussed above are concretely realized in a widely-used open social media platform.

\begin{figure*}[t]
    \centering
    \includegraphics[width=1\textwidth]{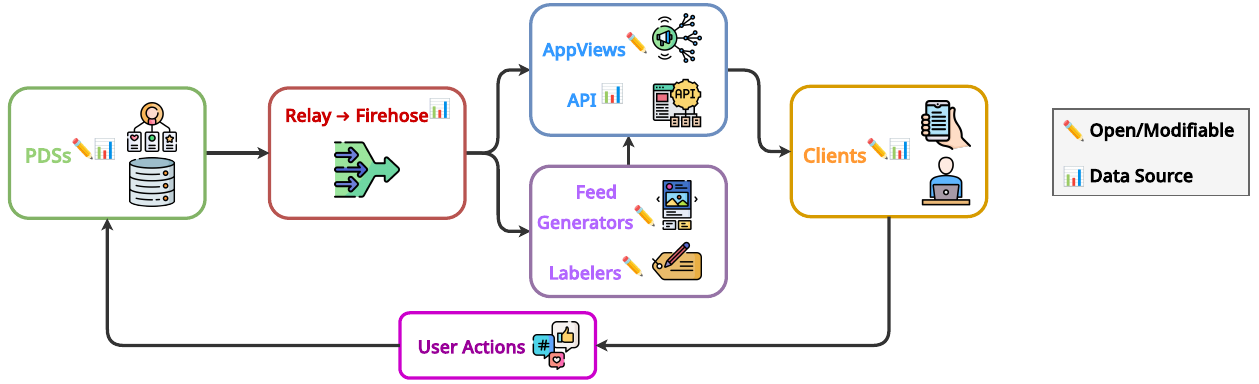} 
    \caption{Schematic diagram of the AT Protocol architecture, with modifiable components and data access infrastructure highlighted. This representation is a simplified adaptation of the architecture presented in the original paper~\citep{kleppmannBlueskyProtocolUsable2024}. We refer readers to that work for more comprehensive technical details. Icons from Flaticon.com.}
    \Description{A flowchart illustrating the AT Protocol architecture and highlighting modifiable components and data access infrastructure. The cycle begins with Personal Data Servers (PDSs), marked as both modifiable and data access infrastructure. Updates from the PDSs flow to the Relay and Firehose, which can act as data access infrastructure. From there, the data feeds into App Views (modifiable), the API (data access infrastructure), Feed Generators (modifiable), and Labelers (modifiable). Finally, these services connect to Clients (both modifiable and data access infrastructure). User actions performed on the Clients then update the PDSs, completing the continuous architectural loop.}
    \label{fig:atProtocolDiagram}
\end{figure*}

\subsection{\bsky{} Infrastructure for OPFEs}
\label{sec:bluesky-elements}
\bsky{} provides a suitable platform on which to instantiate the conceptual framework previously presented. As one of the largest open social media platforms, it has already attracted considerable scientific interest \citep{salloum2025politics,balduf2025bootstrapping,greenwood2026paper}, while offering a user experience and interaction model that closely resemble those of other mainstream platforms (e.g., X). \bsky{} is built around the \atproto{}~\citep{kleppmannBlueskyProtocolUsable2024}, whose architecture separates the underlying social infrastructure from the applications and services built on top of it. Rather than integrating data storage, social relationships, content curation, and user interfaces into a single monolithic system, the protocol exposes these functionalities through a collection of interoperable architectural elements that can be independently implemented, modified, and redeployed. This architectural decomposition provides the foundation that enables \opfes{} on \bsky{}.  Figure~\ref{fig:atProtocolDiagram} illustrates these architectural elements within the AT Protocol. In what follows, we map them onto the open platform elements, showing how the capabilities discussed in Section~\ref{sec:opfes} can be concretely realized on \bsky{}.

\subsubsection{Data Infrastructure}
The data infrastructure of the \atproto{} is centered around \texttt{Personal Data Servers} (\texttt{PDSs}), which store user identities, content, social relationships, and other protocol records. \texttt{Relays} continuously collect repository updates from the multitude of \texttt{PDSs} distributed across the network and expose them internally to the \atproto{} as a single event stream, providing a scalable mechanism through which other architectural components can observe the evolution of the ecosystem. This event stream is exposed to external consumers through the \texttt{Firehose},\footnote{\url{https://bsky.network/docs/consuming-the-firehose/}} or its simplified \texttt{Jetstream} websocket.\footnote{\url{https://bsky.network/docs/jetstream/}} In parallel, indexed views of the same underlying data are made available through \texttt{HTTP APIs}, enabling efficient retrieval of historical information.\footnote{\url{https://endpoints.bsky.app}}

From the perspective of \opfes{}, this infrastructure provides researchers with extensive visibility into platform activity. First, the combination of \texttt{HTTP APIs} and real-time event streams enables the seamless integration of retrospective analyses with live observations \citep{quelle2025academics,pal2026hidden}. Then, these data interfaces allow researchers to observe the complete protocol event stream, rather than only the data associated with their own experiment. Furthermore, unlike conventional platform APIs, these interfaces provide broad, well-documented, and independent access to platform data without requiring discretionary agreements or commercial data-access arrangements. The information exposed through these interfaces is also highly heterogeneous. Besides user-generated content and behavioral events---such as posts, likes, reposts, follows, profile updates, and other interaction traces---the protocol also exposes metadata describing its own architectural components. Protocol services such as \texttt{feed generators} and \texttt{labelers} are themselves represented through openly accessible records and service declarations, making them not only \textit{programmable}, but also \textit{discoverable}, \textit{queryable}, and \textit{analyzable} components of the ecosystem. Collectively, these capabilities instantiate the \textit{Data} element of open platforms, enabling the study not only of user behavior, but also of the very components that influence such behavior~\citep{pedreschi2025human}.

Beyond observation, the protocol also promotes a degree of data sovereignty that is typically impossible on closed platforms. Because the implementation and hosting of \texttt{PDSs} is itself released as open-source software,\footnote{\url{https://github.com/bluesky-social/pds}} researchers may deploy their own instances when needed, thereby controlling how experimental data are collected, stored, shared, and deleted. This prevents experimental data from being permanently incorporated into platform-owned infrastructures. More broadly, the \atproto{} decouples user identity and social data from individual applications, allowing users to migrate between \texttt{PDSs} or \texttt{clients} while preserving their identity and content. From the perspective of social media experimentation, this architectural choice reduces the dependence of experimental infrastructures on a single application or service provider, possibly facilitating longitudinal studies and the preservation of experimental environments even as individual applications evolve or disappear. Overall, the separation between experimental data management and platform operation provides researchers with substantial flexibility in designing data collection pipelines while simultaneously simplifying compliance with data governance and privacy requirements.

\subsubsection{Clients}
\texttt{Clients} are the primary user-facing component of \bsky{}, and represent the applications through which users browse content and interact with the platform. Rather than communicating with a centralized backend, clients interface with the \atproto{} through \texttt{AppViews}, which aggregate and index protocol data to expose application-oriented views. Through these interfaces, clients retrieve social data, render feeds, submit user actions, and interact with other protocol services such as \texttt{feed generators} and \texttt{labelers}. The official Bluesky client, publicly available as open-source software for both web and mobile platforms,\footnote{\url{https://github.com/bluesky-social/social-app}} is only one possible implementation of the user interface. Because the protocol is open, alternative clients can be independently developed and distributed while remaining interoperable with the same underlying social network. Switching clients does not require migrating accounts or rebuilding one's social network, since identity and social relationships are managed independently of the client. This architectural decoupling has fostered a rich ecosystem of community-developed clients and applications.\footnote{\url{https://blueskydirectory.com/clients}} While many provide simple alternative web or mobile interfaces to the platform,\footnote{\url{https://deck.blue/}} others fundamentally rethink how users interact with it.\footnote{\url{https://anisota.net/}} Although these applications differ substantially in interface design, user experience, interaction paradigms, and supported features, they all operate over the same shared data, social graph, and protocol.

From the perspective of \opfes{}, this separation between the underlying social infrastructure and the client provides researchers with a flexible intervention surface. Rather than forcing participants into isolated or synthetic environments, researchers can modify an existing client or develop an entirely new one to experimentally alter how users experience the platform while remaining embedded within a live social media ecosystem \citep{chujo2026exploring}. \bsky{} \texttt{clients} therefore instantiate the \textit{Client and Interface} component of the \textit{Modifiable Components}.

\subsubsection{Feed Generators and Labelers}
Beyond the user-facing \texttt{client}, the \atproto{} exposes two additional backend services that directly influence users' exposure to content: \texttt{feed generators} and \texttt{labelers}. \texttt{Feed generators} are external services responsible for selecting, ranking, and recommending content, allowing different recommendation and data curation algorithms to coexist over the same content.\footnote{\url{https://atproto.com/guides/custom-feed-tutorial}} \texttt{Feed generators} thus determine what content is shown. They do not own the content itself, nor actually produce it---users do. By default, \bsky{} provides users with a small set of built-in feeds, such as the algorithmic \textit{Discover} feed and the chronological \textit{Following} feed. However, researchers and developers can independently create and deploy new feed-generation algorithms~\citep{bradyRedesigningAlgorithms2026, popowski2026social, el2026bonsai, greenwood2026paper}. Users may simultaneously subscribe to an arbitrary number of feeds, each implementing its own curation logic, and seamlessly switch between them within the \texttt{client}. Complementing this mechanism, \texttt{labelers} are independent moderation services that annotate users or content with contextual information and actionable labels.\footnote{\url{https://atproto.com/guides/moderation}} The official \bsky{} moderation service is enabled by default for all users, providing labels such as ``spam,'' ``gore,'' and ``nudity,'' while additional third-party \texttt{labelers}\footnote{\url{https://www.bluesky-labelers.io/}} can be subscribed to concurrently. Therefore, rather than directly filtering content themselves, \texttt{labelers} publish tags that \texttt{clients} interpret according to user-configurable preferences. Users may thus combine labels originating from multiple moderation services and independently configure how each label should be handled, for example by hiding or warning about content flagged with a given label. These services illustrate that in the \atproto{} architecture, content curation and moderation are not exclusively controlled by the platform provider. Instead, they can be re-implemented by independent actors through open components. Multiple \texttt{feed generators} and \texttt{labelers} may therefore coexist, allowing users to customize how content is ranked and moderated without changing the underlying platform, social network, or data.

These open components complement \bsky{} \texttt{clients} and complete the experimental intervention surface. All three components naturally support the construction of experimental counterfactuals, as default and modified implementations can coexist. However, while alternative \texttt{clients} are typically adopted by different groups of users, multiple \texttt{feed generators} and \texttt{labelers} may also coexist within the experience of the same user or group, thanks to simultaneous subscriptions. Collectively, \texttt{feed generators} and \texttt{labelers} respectively instantiate the \textit{Feed and Recommendation} and \textit{Labeling and Moderation} components of the \textit{Modifiable Components}, providing unprecedented opportunities to experimentally study algorithmic curation and platform governance under real-world conditions.

\subsubsection{Governance Infrastructure}
The open platform \textit{Policy and Governance} element is realized in the \bsky{} ecosystem through a collection of publicly available documents governing platform usage, software development, and protocol implementation. Researchers conducting \opfes{} should consult the documents corresponding to the specific platform and protocol components they intend to use or modify. At the platform level, among other artifacts, the Terms of Service and Community Guidelines define the general conditions governing user accounts, content, and acceptable behavior.\footnote{\url{https://bsky.social/about/support/tos}} In addition, researchers accessing platform data through the public APIs should also comply with the Developer Guidelines\footnote{\url{https://bsky.network/docs/developer-guidelines/}} and API documentation,\footnote{\url{https://bsky.network/docs/bluesky-api}} which specify the permitted uses of the available interfaces, technical constraints, and operational requirements. When modifying protocol components, additional governance artifacts become relevant, including the software licenses associated with the corresponding repositories\footnote{\url{https://github.com/bluesky-social}} and the \bsky{} Copyright and Intellectual Property policies,\footnote{\url{https://bsky.social/about/support/copyright}} which regulate the use of the \bsky{} name, logos, and visual identity. 

The practical implications of these governance requirements become particularly evident when modifying existing platform components. For example, although the official \bsky{} client is released as open-source software and can therefore be freely ``forked'' and extended, the accompanying guidelines require developers to perform a number of additional modifications before redistributing the resulting application.\footnote{\url{https://github.com/bluesky-social/social-app}} These include removing or replacing \bsky{} branding and visual identity, updating support and legal references (e.g., Terms of Service and Privacy Policy), and replacing telemetry, analytics, and error-reporting services so that operational data are no longer transmitted to \bsky{}'s infrastructure. Similar practical requirements arise when deploying other protocol components or interacting with platform services. Consequently, implementing an \opfe{} entails considerably more than modifying source code, let alone only the minimal changes required to implement the experimental treatment. Researchers must additionally transform the modified component into an independently deployable system by adapting its supporting software---for example, by replacing telemetry and analytics services---and by establishing the technical, legal, and operational infrastructure required to deploy, maintain, and support the experiment for its duration.

These requirements illustrate a recurring theme of \opfes{}. Openness substantially lowers the institutional barriers to experimentation, but it simultaneously transfers some operational responsibilities from the platform provider to the researchers conducting the experiment.

\subsection{Designing \opfes{} on \bsky{}}
\label{sec:bluesky-example}
We now illustrate how elements of \bsky{} and the \atproto{} can be combined to design an \opfe{}. We revisit the running example introduced in Section~\ref{sec:design-space-approaches}, which investigates the effect of misinformation warning labels on users' propensity to engage with and share content, and map it onto the concrete components presented in Section~\ref{sec:bluesky-elements}. The discussion follows the complete experimental lifecycle, highlighting the practical decisions researchers would need to make, the architectural components involved at each stage, and the trade-offs that arise~\cite{warburton2026conduct}. To make these considerations concrete, we fix one illustrative implementation and follow it throughout the remainder of this section. The specific design choices are not intended to prescribe a canonical or optimal implementation, but rather to demonstrate how the architectural components introduced above can be combined into a coherent implementation blueprint for an \opfe{}.

\subsubsection{Planning}
The research question is fixed by the running example: whether misinformation warning labels affect users’ propensity to engage with or share content. The researchers first decide the experimental design, including the treatment and control conditions, the unit of randomization, the primary outcome variables, and the planned study duration. They also carry out a power analysis to determine their target sample size. They then decide which open platform components to modify. In this example, they opt for an implementation that combines a custom \texttt{labeler} that consumes the \texttt{Firehose} to identify misinformation posts, with a modified \texttt{client} that renders those labels differently depending on the assigned experimental condition. They also decide to extend the \texttt{client} to record fine-grained telemetry, including impressions, dwell time, scrolling behavior, clicks, attempted reposts, and exposure to labeled content. In doing so, they determine which data will be collected, where participant data will be stored, and how long it will be retained. To reduce the implementation burden, they decide to rely on an existing \texttt{PDS} infrastructure for participant data.

Before implementing the modifications, the researchers evaluate their design against the relevant platform documentation, including the Terms of Service, Community Guidelines, Developer Guidelines, branding policies, and the \atproto{} specifications. They further ensure compliance with applicable laws and other ethical requirements. Finally, they submit the protocol for IRB approval and pre-register the study, thereby fixing the experimental design before recruiting participants or collecting data.

\subsubsection{Implementation and Deployment}
The researchers implement the experimental infrastructure. They fork the official \bsky{} \texttt{client} repository and modify it to render misinformation labels according to the assigned treatment condition, while also recording the telemetry defined during the planning phase. As part of this process, they rebrand the application, replace the official support and privacy information with project-specific documentation, and detach services such as crash reporting, analytics, and telemetry collection from the official Bluesky infrastructure, replacing them with infrastructure under their own control. In parallel, they implement and deploy their custom \texttt{labeler}, which continuously consumes the \texttt{Firehose}, identifies misinformation posts according to the selected detection strategy, and publishes the corresponding labels through the protocol. They subsequently deploy both components on their infrastructure, ensuring that the services remain available for the planned duration of the experiment, are adequately secured, and can accommodate the expected participant load.

At this stage, the researchers introduce additional modifications to satisfy further experimental and ethical requirements. They modify the \texttt{client} so that reposts of misinformation are only simulated locally rather than propagated through \bsky{}, allowing participants to experience the interaction while preventing dissemination of harmful content. Similarly, they prevent new posts authored by participants from becoming visible to users outside the study, thereby avoiding the involvement of users who do not participate in the experiment and did not consent to it. Such modifications, however, require carefully balancing experimental \prop{controllability} and \prop{compliance} against \prop{ecological validity}, \prop{experimental isolation}, and \prop{operational accessibility}.

\subsubsection{Recruitment}
The researchers next recruit participants to the experiment. They first decide whether to rely on organic adoption or actively recruit them through channels such as crowdsourcing platforms and social media advertisements. In this example, they opt for active recruitment to obtain the desired sample size within the planned study duration. This choice increases recruitment efficiency, but will likely reduce \prop{ecological validity}, as participants may be unfamiliar with \bsky{} and scarcely represent its native user base. Before granting access to the experimental \texttt{client}, the researchers conduct an onboarding procedure. Participants receive an information sheet describing the study, provide informed consent, complete an initial questionnaire, and are instructed on how to install and use the modified application. Participants who do not already have a \bsky{} account must create one before joining the experiment, introducing an additional source of friction that may affect recruitment and the dropout rate.

\subsubsection{Measurement}
Once the experiment begins, the researchers continuously collect outcome variables through the measurement channels established during implementation. They periodically query the \texttt{HTTP APIs} to retrieve protocol-level information about participant activity, including posts, likes, reposts, follows, replies, profile updates, and other publicly available records. In parallel, the modified \texttt{client} uploads the fine-grained telemetry collected during the study, providing information that is not observable through the \atproto{} alone. These measurement channels serve a dual purpose. Before deployment, the researchers use them to test the modified components, validate the intervention logic, calibrate telemetry collection, stress-test the infrastructure, and verify that treatment assignment is correctly implemented. Once the experiment is underway, the same channels provide a synchronized, multidimensional record of participant behavior throughout the deployment window, combining protocol-level activity with client-side observations to support the subsequent behavioral analysis.

\subsubsection{Completion}
Once data collection concludes, the researchers analyze the outcomes by comparing the treatment and control conditions according to the pre-registered analysis plan. They subsequently offboard participants by debriefing them about the purpose and outcome of the study, resolving any forms of deception, providing access to their data when appropriate, and informing them of any subsequent data retention or deletion procedures.

The researchers then decide how to conclude the experimental deployment. They may decommission the modified \texttt{client}, \texttt{labeler}, and any supporting infrastructure, reducing operational costs and simplifying long-term maintenance. Alternatively, they may continue operating some or all of these components to support longitudinal studies, collect additional data, or make them available as community resources within the \atproto{} ecosystem. This choice therefore balances operational simplicity against the potential scientific and community value of maintaining the deployed infrastructure. Finally, the researchers publish the modified source code, pre-registration report, and, when appropriate, the collected datasets and analysis scripts to facilitate reproducibility.

\section{Discussion}
\label{sec:discussion}
Open and decentralized social media are becoming an increasingly important component of the digital ecosystem~\citep{he2023flocking,balduf2024looking}. Although they have yet to match the massive scale of legacy tech giants, they embody an alternative vision of platform governance centered on openness, transparency, interoperability, and community participation~\citep{kleppmannBlueskyProtocolUsable2024}. Among the many implications of this broader transformation, our work investigates a methodological one: the opportunities these platforms create for conducting independent and ecologically grounded experiments on social media. To this end, we first proposed a design space of social media experimentation, providing a common methodological framework through which heterogeneous experiment archetypes can be systematically positioned and compared. This addresses a gap in the literature by introducing a shared vocabulary and qualitative evaluation framework that clarifies methodological trade-offs and helps researchers align experimental approaches with their scientific objectives. Building on this framework, we showed that \opfes{} occupy a distinct and advantageous region of the design space. Finally, we instantiated \opfes{} on \bsky{} and the AT Protocol, translating the conceptual framework into a concrete open social media ecosystem and providing researchers with a practical blueprint for designing and deploying their own experiments on open platforms. Taken together, these contributions progress from conceptual modeling to platform-specific design. They provide both a theoretical lens for understanding social media experimentation and a concrete pathway for conducting it on open platforms.

\subsection{Temporal Stability of the Design Space}
\label{sec:discussion-temporal}
In formalizing the design space, we sought to identify experimental archetypes and desirable properties that are sufficiently general to remain meaningful as social media research and technologies evolve~\citep{zhang2024form}. While future work may introduce new or hybrid approaches, we expect these to naturally occupy positions within the proposed space rather than fundamentally altering its structure. The framework is therefore intended to provide a stable basis for comparing both existing and emerging forms of social media experimentation.

By contrast, the position of individual experimental archetypes within the design space is inherently dynamic. The degree to which an approach supports a given property depends on a continuously evolving technological, regulatory, and social landscape, including changes in platform architectures, legislation, and governance; ethical standards; and community-developed tools~\citep{bruns2021after,goanta2026great,matias2018civilservant}. Consequently, the scores assigned in our comparative analysis should be interpreted as a characterization of the state of the field at the time of writing, rather than immutable assessments. The \prop{operational accessibility} of \opfes{} illustrates this distinction. We currently assess this property as relatively low, due to the substantial engineering effort required to modify and deploy open platform components. However, as the open social media ecosystem matures and reusable infrastructures and methodological practices become increasingly available~\citep{greenwood2026paper, bradyRedesigningAlgorithms2026}, these operational barriers are bound to diminish, further strengthening the practical viability of \opfes{}.

\subsection{Hybridization of Experimental Approaches}
\label{sec:discussion-hybrid}
We intentionally restricted our analysis to the principal families of social media experiments, without explicitly modeling \textit{hybrid} approaches. This choice kept the design space parsimonious while capturing the methodological foundations from which more complex experimental designs can be constructed~\citep{maclean2020questions}. Nevertheless, hybrid approaches already exist and represent an important direction for future work. The structural trade-offs identified by our comparative analysis naturally motivate the combination of different experimental archetypes, with the goal of jointly exploiting their respective strengths. Existing examples include hybrid laboratory--field experiments, where interventions are deployed on-platform and evaluated partly through post-experiment surveys~\citep{levy2021social, mosleh2022field}.
Additionally, advances in LLM-based social media simulations enable experiments combining simulated and human participants within the same study~\citep{donkers2025understanding}. As the methodological landscape continues to evolve, additional hybridizations are likely to emerge. Furthermore, although \opfes{} are still in their infancy, they are equally amenable to hybridization. Their openness and modularity naturally allow individual platform components and experimental designs to be reused across paradigms. For instance, researchers conducting controlled-environment experiments may reuse a fork of the official \bsky{} client to provide a realistic experimental interface, while client-side studies may leverage \bsky{}'s open-source implementation to simplify the development of browser extensions or experimental clients.
\subsection{Multidimensional Openness, Modularity, and Interoperability}
\label{sec:discussion-openness}
For practical reasons, discussions of social media platforms often contrast \textit{open} and \textit{closed} platforms as two sharply distinct categories. However, openness is not a binary property but rather a multidimensional continuum~\citep{ghazawneh2013balancing}. In practice, platforms differ not only in which components they expose, but also in how those components can be accessed and used~\citep{kleppmannBlueskyProtocolUsable2024}. Consequently, even platforms that are generally regarded as closed may expose specific data or functionalities that enable valuable forms of research. This diversity is well illustrated by X and Reddit, two widely studied platforms. Despite being generally perceived as a closed platform---especially after restricting its once open data APIs~\citep{tromble2021have,bruns2021after}---X has recently open-sourced its recommendation algorithm.\footnote{\url{https://github.com/xai-org/x-algorithm}} Furthermore, it publicly released all data underlying Community Notes, a paramount form of community moderation.\footnote{\url{https://communitynotes.x.com/guide/en/under-the-hood/download-data}} On the contrary, Reddit provided broad access to platform data through APIs and third-party services~\citep{baumgartner2020pushshift} while keeping its core platform components proprietary. These different forms of openness support different forms of scientific inquiry. For example, researchers can inspect and analyze X's recommendation algorithm or study user behavior on Reddit at scale. Nonetheless, neither platform enables the kind of direct intervention that characterizes \opfes{}. Unlike on \bsky{}, researchers cannot modify specific components and deploy the resulting implementations within the platform itself.

This distinction highlights that openness alone is insufficient to support \opfes{}. Equally important are characteristics such as \textit{modularity} and \textit{interoperability}, that is, the ability to independently modify individual platform components while seamlessly reintegrating them into a shared operational ecosystem~\citep{ghazawneh2013balancing}. A platform may expose source code, datasets, or individual services without enabling researchers to compose them into functioning experimental infrastructures. It is precisely the combination of multidimensional openness, modularity, and interoperability that enables the forms of experimentation discussed in this work.

\subsection{Advanced Experiments}
\label{sec:discussion-pushing}
Up to this point, we have examined the properties of \opfes{} from the perspective of an individual researcher or a small research group. However, when greater technical capabilities are available---such as when operating within a larger organization or community, or when leveraging a rich library of reusable software artifacts---a key limitation of \opfes{} can be effectively offset, paving the way for more advanced experimental setups.
In subsection \ref{sec:opfes-design-space}, we rated the \prop{controllability} of \opfes{} with a medium score, because the property is bounded by the availability of existing open components. However, this limitation can be overcome if researchers possess the capacity to develop the missing elements themselves. For example, an entire subsystem of a social platform can be swapped out---enabling custom user interfaces built on shared underlying protocols and code, or entirely new recommendation and moderation algorithms.
What follows are a few scenarios illustrating such advanced experiments.

\paragraph{Experimental User Interfaces}
There is growing interest within Human-Computer Interaction (HCI) in studying social media UI mechanisms---notably infinite scroll and push notifications~\citep{ruiz2024design, degala2026doom, monge2023nudging, herder2024defeat, gilbert2025101}. This focus stems from their well-documented impacts on memory~\citep{ruiz2024design, degala2026doom}, attention, ability for the user to disengage~\citep{gilbert2025101}, and overall digital well-being. Existing studies have explored various mitigation strategies, ranging from behavioral nudging and warning prompts to radical UI redesigns. Many of these intervention concepts would benefit from replication as \opfes{} to gather field evidence under conditions of higher ecological validity and at a significantly larger scale of user participation.
Naturally, the required technical development effort scales with the depth of the UI modification---ranging from lightweight adjustments of visual parameters (e.g., changing the background lightness as the user scrolls in~\citep{monge2023nudging}) to comprehensive reimplementations of full interface modules (e.g., custom pagination schemes in~\citep{herder2024defeat} or the home screen redesign in~\citep{monge2023nudging}).
Experiments are not limited to addressing problems identified in the literature, but they could also be used to explore alternative UI design strategies. For example, participants in an experiment could interact with live Bluesky content via a Reddit-style long-form blogging interface, offering insights on how such a change in the presentation layer alters user experience. Real-world implementations like deck.blue\footnote{\url{https://deck.blue/}} and RedGazer\footnote{\url{https://www.redgazer.com/}} demonstrate the technical feasibility of alternative Bluesky interfaces, even if they are built for consumer use rather than scientific research.

\paragraph{New Recommendation and Moderation Algorithms} A wide variety of fields, ranging from computational social sciences to political science and psychology, are deeply interested in understanding the individual and collective consequences of social media, both offline and online, and how these consequences are shaped by what users see and under which conditions. These consequences include polarization, toxicity, shaping of political attitudes or mental health issues, among others. This has driven research into how recommendation and moderation algorithms should be developed and deployed, as well as how existing ones contribute to these consequences. In order to adequately address these complex problems, studies require a high level of ecological validity, control, and transparency rarely available for independent researchers. This has led to the most robust research being developed together with platforms to study, for instance, the effects of recommendation algorithms during elections ~\citep{guessHowSocialMedia2023, guessResharesSocialMedia2023} or alternative forms of content moderation~\citep{horta2025post}. However, without these collaborations, promising research directions in the field, such as personalized content moderation~\citep{moscato2025personalization} or recommendation~\citep{jhaver2023personalizing} struggle to fully answer the questions at hand. This limitation extends to a wide range of questions proposed by experts in the field, such as those from initiatives like the Social Data Science Alliance, which has outlined critical questions that could be addressed under Article 40 of the DSA~\citep{knott2025article}. Many of these limitations can be overcome with the combination of greater ecological validity and openness that OPFEs provide, unlocking research until now restricted to platform partnerships or limited deployments. For instance, studies have begun exploring critical questions such as how different recommendation algorithms influence what users see during national elections on Bluesky~\citep{bradyRedesigningAlgorithms2026}. Similarly to experimental user interfaces, alternative recommendation and moderation algorithms are being developed for consumer use with and without scientific roots. Examples include MySky,\footnote{\url{https://www.mysky.social/p/introducing-greenearth}} which aims to create both prosocial~\citep{stray2026prosocial} feeds and open recommender architectures, with strong attention to the \atproto{}.

\paragraph{Novel Dynamics} At a higher level of complexity, researchers can also experiment with dynamics unique to the open social web, where disparate applications and media types interoperate across a shared underlying protocol. An example on the AT Protocol is the Standard Site schema\footnote{\url{https://standard.site/}} for long-form publishing and related applications (e.g., Standard Reader,\footnote{\url{https://standard-reader.app/}} Leaflet,\footnote{\url{https://leaflet.pub/}} Pckt\footnote{\url{https://pckt.blog/}} or Offprint\footnote{\url{https://offprint.app}})
By working with this environment, researchers can conduct experiments on how users interact with long-form content \emph{embedded} in a social media context focused on microblogging. Bluesky lacks native support for long-form content in its UI, but seamlessly delegates these content types to third-party, protocol-compatible applications---such as those relying on Standard Site---without requiring users to log in with a different provider. At the time of writing, this novel approach for interacting with content has yet to prove itself, providing a fertile ground for research.

\subsection{Closing the Feedback Loop Between Research and Practice}
\label{sec:discussion-feedback-loop}
The opportunities created by open platforms extend well beyond enabling better social media experiments. Historically, social media research and platform development have evolved largely independently. Researchers could observe and critique platform behavior, but rarely had the agency to translate findings into deployed changes on live systems~\citep{freelon2018computational,bruns2021after}. Consequently, platform improvements depended on discretionary adoption by platform providers, leaving a persistent gap between research and practice. Open platforms may fundamentally shorten this feedback loop.

\paragraph{Developers and Platform Administrators.} For platform developers, this architecture facilitates the direct incorporation of research outcomes into production systems. Researchers can contribute improvements to existing components through established open-source development practices, or release alternative implementations that remain interoperable with the broader ecosystem~\citep{greenwood2026paper}. Consequently, experimental findings need not remain confined to academic publications, but can instead materialize as deployable improvements. The traditional separation between evaluating a platform and improving it is therefore reduced.

\paragraph{Platform users.} The same mechanisms also benefit end users. Experimental components that prove effective may continue to exist beyond the lifetime of a study, allowing users to adopt improved platform components~\citep{greenwood2026paper}. Even when multiple implementations coexist without a single objectively superior solution, the resulting diversity expands user choice and enables individuals to tailor their social media experience to their own preferences and needs~\citep{zhang2024form}. In this sense, research contributes not only to scientific understanding, but also to enriching the broader ecosystem of tools available to platform communities. These dynamics may create a virtuous feedback loop, where a richer ecosystem attracts additional users, whose participation further expands the ecosystem~\citep{la2021understanding,quelle2025academics} and, in turn, creates new opportunities for platform evolution and scientific research.

\paragraph{Researchers.} Further long-term implications concern the research community itself. Unlike traditional one-off experimental systems, the software artifacts produced as part of a research project can become reusable building blocks for subsequent studies. As research artifacts accumulate, experimentation could become increasingly cumulative, with each project contributing infrastructure that subsequent studies can directly reuse rather than reimplement from scratch~\citep{greenwood2026paper, bradyRedesigningAlgorithms2026}. For example, a research team may release a modified \bsky{} client specifically designed for scientific data collection, allowing users to explicitly opt into continuous telemetry donation while automatically handling compliant data management (e.g., consent, anonymization, secure data transfer). Compared to contemporary data donation workflows that typically put a hefty burden on both donors and researchers~\citep{carriere2025best}, such a research-oriented client could seamlessly collect the needed measurements. Similarly, a different reusable client could expose configurable intervention mechanisms that allow researchers with limited technical expertise to deploy common classes of experiments without the need to directly modify the underlying software. Users choosing to adopt such a client could consent to participate in multiple studies over time, allowing independent research teams to reuse a common experimental infrastructure while preserving transparency and informed consent. Looking further ahead, these developments may enable new forms of citizen science, transforming social media platforms from passive objects of scientific observation into active infrastructures for collaborative research. This ambitious vision for research ecosystems would likely be sustained by non-profit support structures that incubate open, public-interest projects about the AT Protocol and social technologies in general, like the Modal Foundation.\footnote{\url{https://www.modalfoundation.org/}}
In conclusion, as anticipated in Section~\ref{sec:discussion-temporal}, as shared research infrastructures and communities mature, the \prop{operational accessibility} of \opfes{}---currently one of their principal limitations---may improve substantially.

\subsection{Scientific Sovereignty Through Open Platforms}
In recent years, open and decentralized platforms have attracted increasing attention in discussions surrounding digital and technological sovereignty~\citep{burwell2026digital}. These debates have primarily focused on reducing the concentration of digital infrastructures under a small number of private actors and on enabling more transparent, participatory, and inclusive online ecosystems. Following the previous discussion of the long-term research implications of open platforms, our work highlights an additional consequence of this transformation: the emergence of open platforms as scientific infrastructures. Historically, public researchers have largely depended on proprietary platforms to validate hypotheses~\citep{freelon2018computational,bruns2021after}. Consequently, the scientific agenda has often been shaped by the degree to which platforms were willing to provide researchers with access to data, APIs, or experimental collaborations~\citep{murthy2024sociology,goanta2026great, tromble2021have}. By contrast, open platforms shift part of this agency from platform operators to the research community itself. In this sense, openness promotes a form of \emph{scientific sovereignty}: the ability of the scientific community to independently generate, validate, and reproduce evidence about digital platforms without relying exclusively on discretionary collaboration with their operators.

\subsection{Community Values and the Evolution of Open Platforms}
Throughout this work we have discussed the experimental opportunities created by open platforms primarily in terms of their technical architecture and governance. However, these opportunities are equally shaped by the values and practices of the \textit{communities} that build, maintain, and inhabit these ecosystems~\citep{muralidharan2026federating,sokoto2026open}. The experimental capabilities of \opfes{} should therefore not be regarded as immutable properties of individual platforms, but as emergent properties of broader socio-technical ecosystems. Open platforms are frequently contrasted with traditional closed platforms, yet openness itself is neither static nor guaranteed. As these ecosystems grow, changes in user base, governance, commercialization, and infrastructure costs may alter the incentives that originally motivated their design~\citep{sadowski2008transition}. Consequently, the degree of openness, modularity, interoperability, or community participation exhibited by a platform may evolve over time. Adopting an open protocol does not necessarily imply embracing the collaborative norms traditionally associated with the surrounding community, as illustrated by the recent controversy\footnote{\url{https://standard-reader.app/a/did:plc:s2rczyxit2v5vzedxqs326ri/3movp4n7tgc2e?q=0O_RBR2m5t}} following the launch of \textit{W Social} as part of the AT Protocol ecosystem.\footnote{\url{https://blog.elenarossini.com/w-social-uncovered-the-reality-behind-the-hype/}}
Similarly, technical openness alone is insufficient to sustain the experimental opportunities discussed in this paper. These opportunities also depend on a culture of reciprocity in which researchers, developers, and users actively contribute software, infrastructure, documentation, and methodological innovations back to the ecosystem. Emerging community initiatives within the \atproto ecosystem, such as the ATmosphereConf,\footnote{\url{https://atmosphereconf.org/}} exemplify the active cultivation of these collaborative norms. Communities are also at the forefront in tackling a fundamental criticality.
Currently, platform openness depends on a private entity, \textit{Bluesky Social PBC}, maintaining the servers and further developing the AT Protocol. Most of this work is publicly released, so a third party could in principle take over. This has in fact already been put into practice to various degrees of independence with the development of \textit{Blacksky},\footnote{\url{https://blackskyweb.xyz/}} \textit{Eurosky},\footnote{\url{https://eurosky.tech/about/}} and \textit{LeafPlaza},\footnote{\url{https://www.leafplaza.eu/}} among other projects. Such continuity of support is precisely what the sustainability of OPFEs requires, since these depend, at a minimum, on open platforms with a sufficiently large user base.

\subsection{Legal and Ethical Considerations}
Conducting ecologically valid field experiments on online platforms has become one of the central methodological and ethical challenges of contemporary computational social science~\citep{polonioli2023ethics}. The closer a study comes to observing or manipulating real-world behavior under natural conditions, the more likely it is to interact with multiple layers of ethical, legal, and institutional governance, including human-subject research principles (e.g., informed consent, deception, proportionality), data protection law (e.g., GDPR), platform regulation (e.g., the EU DSA and AI Act), and the contractual obligations imposed through platforms' Terms of Service. Recent controversies illustrate how narrow the margin for error has become. The widely criticized University of Zurich experiment on Reddit,\footnote{\url{https://www.reddit.com/r/changemyview/comments/1k8b2hj/meta_unauthorized_experiment_on_cmv_involving}} in which undisclosed AI-generated accounts attempted to persuade users without their knowledge, triggered international debate over the ethical boundaries of online experimentation~\citep{ogrady2025unethical}. More recently, other field experiments employing automated accounts and covert behavioral observation have raised similar questions, despite involving substantially different designs and milder manipulations~\citep{oda2026field}. These examples illustrate the broader challenge of conducting scientifically valuable field experiments while simultaneously satisfying an increasingly complex landscape of ethical, legal, contractual, and societal expectations. Notably, both cited studies were pre-registered and received approval from their respective IRBs. This highlights an important distinction: pre-registration and IRB approval remain essential safeguards for scientific integrity, but neither guarantees full ethical and legal compliance~\citep{polonioli2023ethics}. More fundamentally, these cases expose a growing mismatch between contemporary social media experiments and the existing oversight mechanisms. Social media experiments are no longer evaluated solely through the lens of research ethics, but increasingly intersect with platform policies, contractual obligations, privacy regulation, and emerging AI-specific legislation. Addressing this complexity therefore requires broader models of oversight capable of considering the full regulatory landscape in which social media research is conducted.

From this perspective, \opfes{} present both opportunities and challenges. On one hand, they alleviate some of the tensions that characterize experimentation on closed platforms, supporting greater transparency, reproducibility, compliance, and independent scrutiny. At the same time, however, they also transfer agency from platform providers to researchers. With greater freedom to design, deploy, and operate interventions comes greater responsibility for ensuring that the research remains ethically justified and legally compliant. Freedom to experiment should therefore not be misconstrued as a license for unchecked experimentation. Rather, it heightens the need for responsible practice and rigorous oversight.

\section{Conclusions}
\label{sec:conclusions}
We investigated how open social media platforms give rise to a new class of \textit{open-platform field experiments} (\opfes{}) and how these relate to existing experimental approaches. To this end, we first formalized the design space of social media experimentation by systematizing the main families of existing approaches and identifying the desirable properties through which their capabilities can be characterized. Our comparative analysis revealed the structural trade-offs that have historically shaped this landscape and showed that different approaches occupy distinct regions of the design space. We then introduced \opfes{}, demonstrating that they colonize a previously unexplored region by combining high \prop{ecological validity} with strong support for properties such as \prop{observability}, \prop{compliance}, and, to a lesser extent, \prop{controllability} and \prop{platform independence}. Finally, we mapped this conceptual framework onto \bsky{} and the \atproto{}, illustrating how these capabilities emerge from concrete architectural components and providing a blueprint for designing a complete \opfe{}.

Beyond characterizing a new experimental paradigm, this work provides both conceptual and practical tools for future research. The proposed design space offers a common framework for positioning, comparing, and reasoning about existing and emerging experimental methodologies, while the comparative assessment can support researchers in selecting approaches that best match their objectives, requirements, and resources. The analysis of \opfes{} and their realization on \bsky{} further provides a practical reference for researchers interested in conducting experiments on open social media. More broadly, by highlighting how open platform architectures fundamentally expand the possibilities for independent, transparent, and reproducible causal research, this work contributes to establishing \opfes{} as a practical methodological paradigm and stimulates further experimentation within the growing ecosystem of open social media.

\begin{acks}
This work is partly supported by the European Union with the ERC project DEDUCE under grant \#101113826.
\end{acks}

\bibliographystyle{ACM-Reference-Format}
\bibliography{modsky} 

\clearpage
\appendix

\renewcommand{\thetable}{A.\arabic{table}}
\setcounter{table}{0}

\section{Details of Typology Assessment Scores}
The approach--property scores reported in Table~\ref{tab:approaches-properties} are qualitative and comparative assessments of broad methodological tendencies across families of experimental approaches. Because individual implementations may differ in practice, the scores should not be interpreted as precise or absolute measurements. To increase transparency and clarify the reasoning underlying this comparative analysis, Table~\ref{tab:scores-motivations} reports a brief motivation for each approach--property score assignment.

{\small
\setlength{\tabcolsep}{2pt}
\begin{longtable}{
    @{}
    >{\raggedright\arraybackslash}p{2.6cm}
    @{\hspace{10pt}}
    >{\raggedright\arraybackslash}p{3.2cm}
    >{\centering\arraybackslash}p{1cm}
    >{\raggedright\arraybackslash}p{6.5cm}
    @{}
}

\caption{Motivations for the scores reported in Table~\ref{tab:approaches-properties}. Each entry provides a brief justification for the extent to which a given experimental approach tends to support a specific property.}
\label{tab:scores-motivations} \\
\toprule
\textbf{approach} & \textbf{property} & \textbf{score} & \textbf{motivation} \\
\midrule
\endfirsthead

\multicolumn{4}{l}{\footnotesize\textit{continued from previous page}} \\
\toprule
\textbf{approach} & \textbf{property} & \textbf{score} & \textbf{motivation} \\
\midrule
\endhead

\midrule
\multicolumn{4}{r}{\footnotesize\textit{continues on next page}} \\
\endfoot

\bottomrule
\endlastfoot

\multirow{9}{=}{Survey-based experiments}
    & Ecological validity       & \low  & Survey environment differs from real use and cannot replicate social environment \\ %
    & Controllability           & \high & Stimuli and environment fully controlled by researcher \\
    & Observability             & \med  & Primarily limited to self-reported measures \\
    & Compliance                & \high & IRB and consent easily enforced \\
    & Platform independence     & \high & No platform cooperation needed \\
    & Scalability               & \low  & Recruiting many participants is costly and non-organic \\
    & Operational accessibility & \high & Easy to deploy via established survey tools and procedures \\
    & Experimental isolation    & \high & No confounds from live system \\
    & Reproducibility           & \high & Survey tools support replication \\
\midrule

\multirow{9}{=}{Controlled-environment experiments}
    & Ecological validity       & \low  & Lab setting differs from real use and cannot replicate social environment \\
    & Controllability           & \high & Stimuli and environment fully controlled by researcher \\
    & Observability             & \med  & Behavior observed but not full logs \\
    & Compliance                & \high & IRB and consent easily enforced \\
    & Platform independence     & \high & No platform cooperation needed \\
    & Scalability               & \low  & Recruiting many participants is costly and non-organic \\
    & Operational accessibility & \med  & Requires environment setup \\
    & Experimental isolation    & \high & No confounds from live system \\
    & Reproducibility           & \high & Controlled environments support replication \\
\midrule

\multirow{9}{=}{In situ experiments\\on closed platforms}
    & Ecological validity       & \high & Real users in natural context \\
    & Controllability           & \low  & Platform rules and architecture limit manipulation \\
    & Observability             & \med  & Available data and measures may be limited by platform features \\
    & Compliance                & \low  & Adherence to platform policies and ethics can be challenging \\
    & Platform independence     & \med  &  Restricted by platform affordances and possibilities \\
    & Scalability               & \high & Large user bases available \\
    & Operational accessibility & \med  & Difficulties may arise from existing platform policies and features \\
    & Experimental isolation    & \low  & Confounding effects and noise from live system \\
    & Reproducibility           & \med  & Live system deployment affects replications \\
\midrule

\multirow{9}{=}{Client-side experiments}
    & Ecological validity       & \med  & Users in real context but artificially modified by researchers \\
    & Controllability           & \med  & Limited to client-side possibilities \\
    & Observability             & \med  & Client events captured but other behavioral measures and data opaque \\
    & Compliance                & \med  & Consent and data handling can depend on underlying platform \\
    & Platform independence     & \med  & Limited by platform client-side and data availability \\
    & Scalability               & \low  & Deployment reach is limited and can be costly \\ 
    & Operational accessibility & \low  & Requires client-side tool development and distribution \\
    & Experimental isolation    & \med  & Confounding effects and noise from live system, but possible client-side control \\
    & Reproducibility           & \med  & Client logic replicable but live system varies \\
\midrule

\multirow{9}{=}{Simulation-based experiments}
    & Ecological validity       & \low  & Synthetic agents lack real behavior \\
    & Controllability           & \high & All parameters set by researcher \\
    & Observability             & \high & Full system state observable \\
    & Compliance                & \high & Involves no humans or platform \\
    & Platform independence     & \high & Fully self-contained, no platform needed \\
    & Scalability               & \high & Simulations scale computationally \\
    & Operational accessibility & \low  & High cost and difficulty to build realistic simulations \\
    & Experimental isolation    & \high & Perfect isolation by construction \\
    & Reproducibility           & \high & Simulation reruns easily repeatable \\
\midrule

\multirow{9}{=}{Platform-run experiments}
    & Ecological validity       & \high & Real users in natural context \\
    & Controllability           & \high & Platform has access to system-level manipulations \\
    & Observability             & \high & Platform has access to rich large-scale behavioral measures \\
    & Compliance                & \med  & Inherently compliant with platform rules, but typically lacks explicit user consent \\
    & Platform independence     & \low  & Researcher depends on platform by construction \\
    & Scalability               & \high & Large user bases available \\
    & Operational accessibility & \high & Mostly conducted by the platform itself \\
    & Experimental isolation    & \low  & Difficult control over confounding effects from live system \\
    & Reproducibility           & \low  & New collaboration with platform required and live system difficult to replicate \\
\midrule

\multirow{9}{=}{Open-platform\\field experiments}
    & Ecological validity       & \high & Real users in natural context \\
    & Controllability           & \med  & Researcher can manipulate available open components \\
    & Observability             & \high & Usually available tools to fetch data and observe behavior \\
    & Compliance                & \high & Open protocols support ethical review \\
    & Platform independence     & \med  & Platform terms may apply \\
    & Scalability               & \med  & Usually requires active opt-in and limited number of users in existing open platforms\\
    & Operational accessibility & \low  & Implementation can require high technical knowledge \\
    & Experimental isolation    & \low  & Difficult control over confounding effects from live system \\
    & Reproducibility           & \med  & Live system difficult to replicate but easy to share open components \\

\end{longtable}
}

\end{document}